\documentclass[twocolumn]{aastex631}  
\usepackage{enumitem}
\usepackage{xcolor}
\usepackage{xcolor, soul}
\sethlcolor{yellow}
\usepackage{subfigure}
\usepackage{tablefootnote}
\usepackage{bm}
\usepackage{lineno}
\hypersetup{backref=true,pagebackref=true,  hyperindex=true,colorlinks=blue,breaklinks=true,  urlcolor=violet,linkcolor= red, bookmarks=true, 	bookmarksopen=true,  filecolor=cyan, citecolor=blue, linkbordercolor=blue}
\shorttitle{Self-lensing/Eclipsing signals in Edge-on Double White-Dwarf Systems}
\shortauthors{Sajadian}

\begin{document}
\title{Study Self-lensing/Eclipsing Signals in Edge-on Double White-Dwarf Systems}

\author[0000-0002-0167-3595]{Sedighe Sajadian}
\affiliation{Department of Physics, Isfahan University of Technology, Isfahan 84156-83111, Iran, \url{s.sajadian@iut.ac.ir}}

\begin{abstract}
Stellar lightcurves from edge-on double white dwarf systems(DWDs) have periodic lensing/eclipsing signals at times of alignment between two components as seen by the observer. Here, we study the characterization and detection of these signals. In common DWDs, the Einstein radii have similar orders of magnitude with WDs' radii, and the projected source and lens radii normalized to the Einstein radius ($\rho_{\star}$, and $\rho_{\rm l}$) are $\sim 1$. Both of them are reduced with the orbital period and the lens mass. If $\rho_{\rm l}\simeq 1$ the lensing-induced minor image is always blocked by the lens which results lower magnification factors. If $\rho_{\rm l}\lesssim 1$ and in transit events the finite-lens effects decrease the lightcurves' width. When $\rho_{\rm l}\gtrsim1$ (happens for close DWDs including one low-mass and one massive WD) deep or complete eclipses dominate to lensing effects. The self-lensing signals maximize for massive DWDs in wide orbits. We study the detect-ability of lensing/eclipsing signals in edge-on DWDs in observations by The NASA's Transiting Exoplanet Survey Satellite(TESS), The Vera Rubin Observatory(LSST) and The Nancy Grace Roman Space Telescope. We simulate stellar lightcurves due to edge-on DWDs and generate synthetic data points based on their observing strategies. Detection efficiency maximizes for extremely low-mass WDs in close orbits, and the numbers of DWDs within 100 pc and an observing cone with detectable lensing/eclipsing signals in one $27.4$-day TESS and $62$-day Roman observing window are $\sim1$ and $<1$, respectively. Detecting these signals by LSST is barely possible because of its long cadence.      
\end{abstract}


\section{Introduction}\label{sec2}
Evolutionary tracks of different stars reveal that stars with initial masses lower than $8 M_{\odot}$ will evolve to white dwarfs (WDs). WDs are the compact objects with masses in the range $\sim [0.17-1.4] M_{\odot}$, and average radii $\sim0.01 R_{\odot}$, which decrease with increasing mass \citep[see, e.g., ][]{1972ApJWDMR,2020Federico}. Galactic models predict that around ten billion WDs live in our galaxy, and $9$-$14\%$ of these WDs are double (DWDs, \citealp{2017AAToonen}). However, in some other references the predicted binarity fraction for WDs in our galaxy is $\lesssim 1\%$ \citep[e.g.,  ][]{2001Nelemans}. Mainly, DWD systems are generated from binary stars. The separations of WDs in such systems are predicted to be lower than their initial separations, because when the lower-mass stellar companion is in the red giant phase and overfills the Roche lobe, a mass-transfer and as a result a reduction of their separation happens.  

There are several methods to realize binarity of WDs, depending on their orbital periods, orbital inclination angles, and their surface temperatures. These methods can be classified into (i) spectroscopic manifestation from either radial velocities (through periodic variations of strong absorption lines' location) or their double-lined overlapped luminosity \citep[see, e.g., ][]{2020kilicmnras}, (ii) photometric observations to capture either periodic eclipsing or self-lensing signals \citep{2021GrazaWilson,2023MNRASMunday,2024arXivJin}. However, in close DWDs some other photometric variations could happen such as Doppler boosting \citep{2010ApJShoperer}. Nowadays, there are several telescopes and facilities which are able to captures these signals, such as Gaia, the NASA's Transiting Exoplanet Survey Satellite (TESS, \citealt{Ricker2024}), the Vera C. Robin Observatory which was named as the Large Synoptic Survey Telescope (LSST, \citealt{2009LSSTbook}), the Zwicky Transient Facility (ZTF, \citealt{2019PASPbellem}), etc. \citep[see, e.g., ][]{2023arXivNir,2024sajadianAS}.

Close DWD systems, with short orbital periods from several hours to minutes, are very interesting targets because they are potential sources for low-frequency gravitational waves detectable by space-based interferometers such as the Laser Interferometer Space Antenna (LISA, \citealt{2017AmaroSeoane}), TianQin \citep{2016Luo}, and Taiji \citep{Ruan2020}. In addition, they are progenitors of Type Ia supernovae. PTF J0533$+$0209 is an example of a close DWD system with the orbital period $20$ minutes including a CO WD and an extremely low-mass (ELM) He WD \citep{2019ApJBurdge}. The formation of this target is on debate, e.g., \citet{2022ApJChen} discussed that this target has not likely formed through common-envelope (CE) evolution.

\noindent Concerning the formation of close DWDs, two potential channels exist. In the first channel a donor star is rotating around a CO WD in a close orbit, so that a stable mass-transfer occurs between them. Such systems are converted to DWDs including a CO WD and He WD. For such systems there is a correlation between He WD mass and the orbital period \citep{1999AATauris}. Through this mechanism, ELM WDs can be born \citep{2019ApJLi,2021Soethe}. There is another mechanism which produces close DWD systems, in which a CO WD and red giant have a CE with an unstable mass-transfer. Then and after ejecting CE, a DWD system (CO WD and He WD) will form.

Most of DWDs discovered up to now are close and have short orbital periods. For instance, \citet{2020ApJBrown} found 98 DWDs through the ELM spectroscopic Survey \citep{2010ApJELTsurvey} while all targets had the orbital periods shorter than $1.5$ days. Also, \citet{2024MNRASmunday} found 34 DWD systems with double-line signatures in their spectra with the orbital periods shorter than $13.5$ days. The main reason is that detecting interacting binaries is easier than detecting detached ones, although the number of detached binaries should be higher than interacting ones. For instance, \citet{2004AAwillems} showed that wide and non-interacting binaries including white dwarfs generally would comprise more than $75\%$ of total binary systems containing white dwarfs. If detached DWDs are edge-on as seen by the observer, we can realize their binarity by capturing either their self-lensing or eclipsing signals.

In this paper, we study characterization and detection of gravitational self-lensing signals reduced by eclipse in edge-on DWDs. Self-lensing happens in binary systems including compact objects whose orbital plane is edge-on as seen by the observer \citep{1973AAMaeder,2021MNRASWiktorowicz,2024sajadianAS}. During self-lensing signals due to compact objects some part of the images's area will be blocked by the compact object's disk, which is the so-called finite-lens effect or occultation or eclipse \citep[see, e.g., ][]{2002ApJAgol}. The finite-lens effects decrease or even remove the self-lensing signals, while they are not recognizable from lensing signals in real observations \citep{2016ApJHan,2024sajadianfinite}. The probability of occurring self-lensing signals in binary systems was first estimated by \citet{1995ApJGould,1997ChPhLQin}. \citet{2002AABeskin} roughly estimated the number of self-lensing signals that can be detected by the Sloan Digital Sky Survey (SDSS, \citealt{2000SDSSyork}) due to different binary systems including compact objects through analytical evaluating probability functions. Based on the same method, \citet{2023arXivNir} roughly evaluated the number of detectable self-lensing signals in edge-on binary systems that can be detected by the new generation of ground and space-based telescopes, such as the Vera C. Robin observatory, TESS telescope, etc. Up to now, several self-lensing signals due to binary systems including WDs and main-sequence stars (WDMS) have been discovered through the Kepler and TESS data \citep{KruseAgol2014,2018AJKawahara,2019ApJLMasuda,2024ApJSorebella}.

The outline of the paper is as follows. We explain our formalism to simulate these systems and their self-lensing/eclipsing signals in Section \ref{sec3}. Additionally, in this section we evaluate the lensing parameters, and lensing/eclipsing signals in all possible DWD systems. In the next section, \ref{sec4}, we numerically calculate self-lensing/eclipsing light curves using inverse-ray-shooting (IRS) method and discuss on their properties. In Section \ref{sec5}, we investigate detectability of these signals through three observing strategies by the TESS, LSST and Roman telescopes. Finally, we conclude and review the results in Section \ref{sec6}.  
\begin{figure*}
\centering
\includegraphics[width=0.325\textwidth]{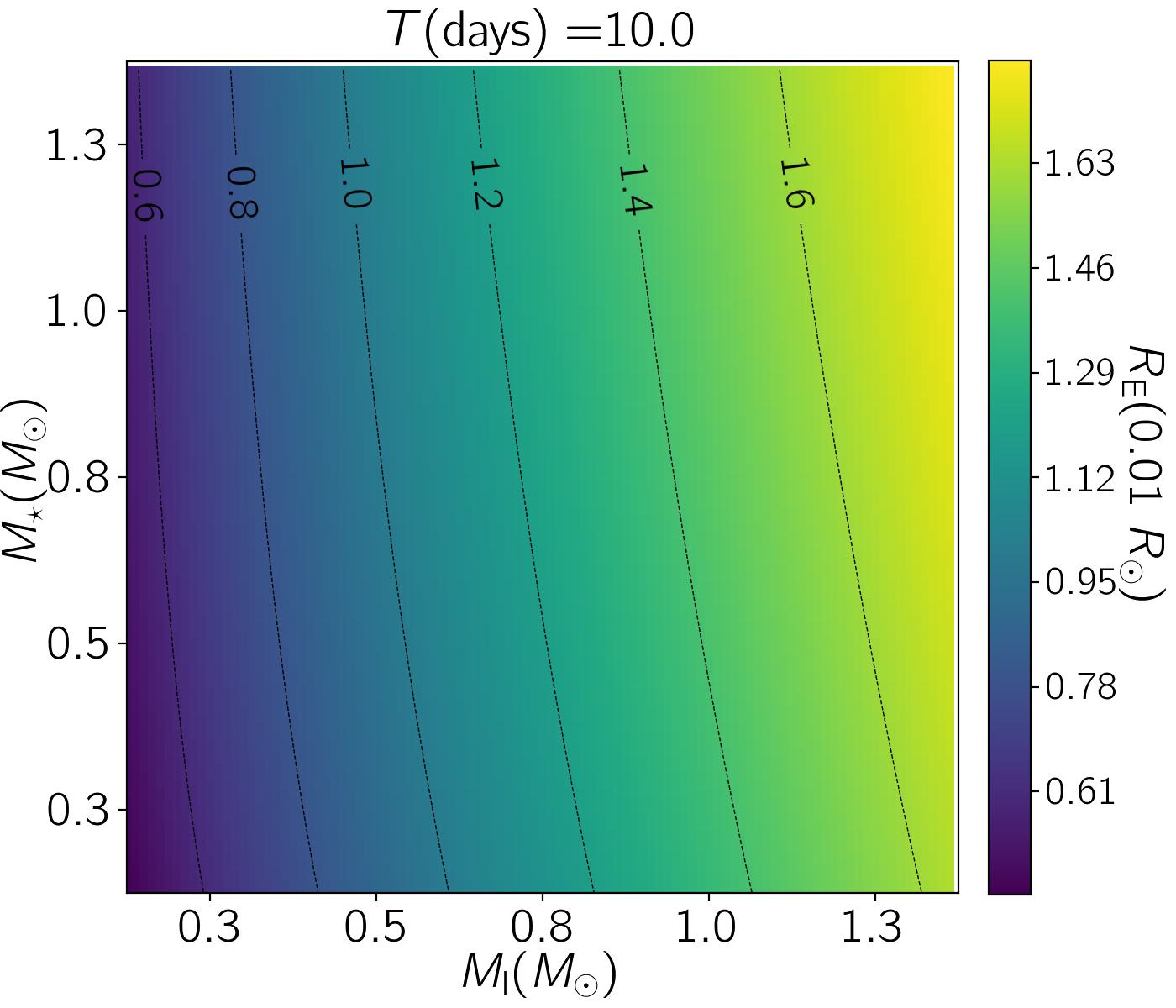}
\includegraphics[width=0.325\textwidth]{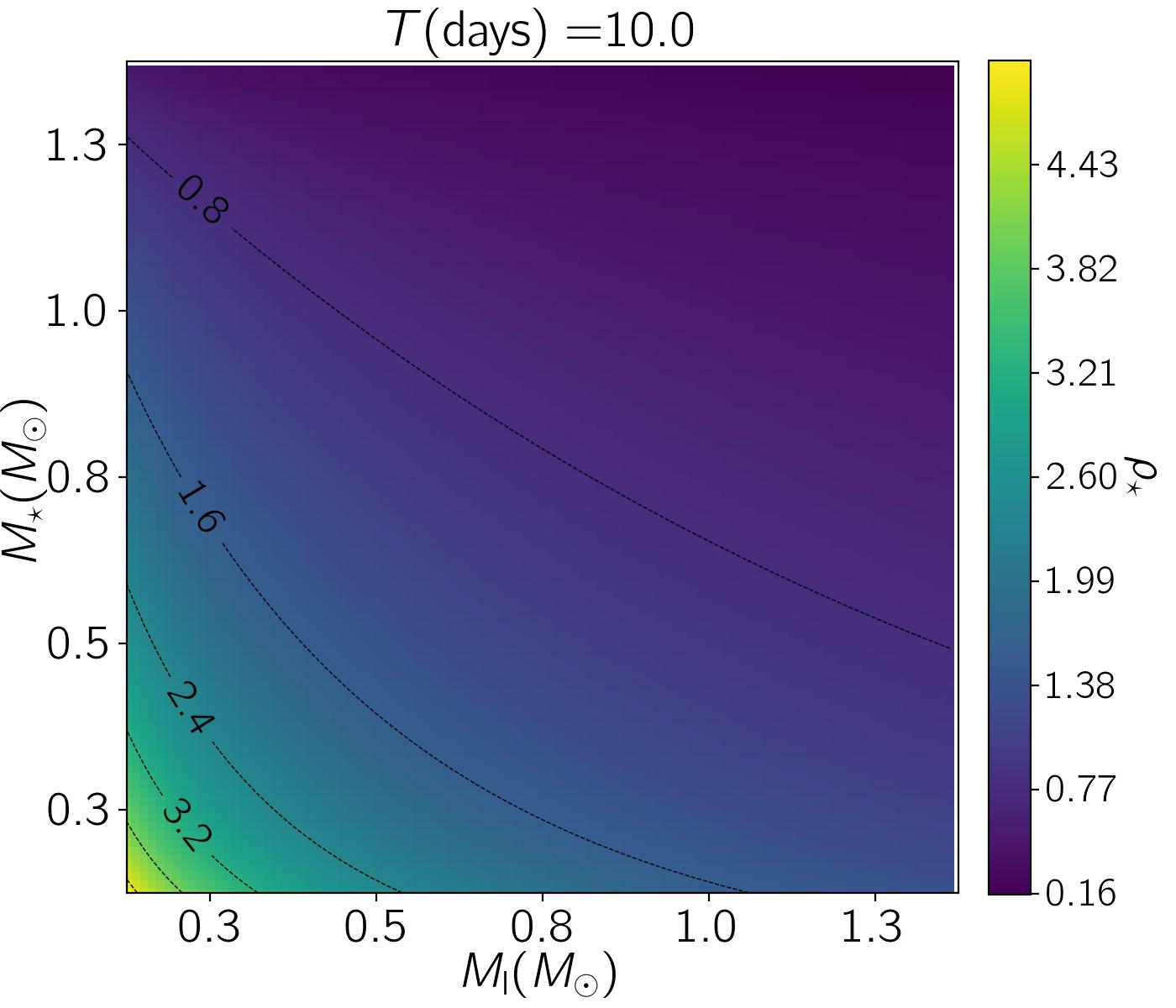}
\includegraphics[width=0.325\textwidth]{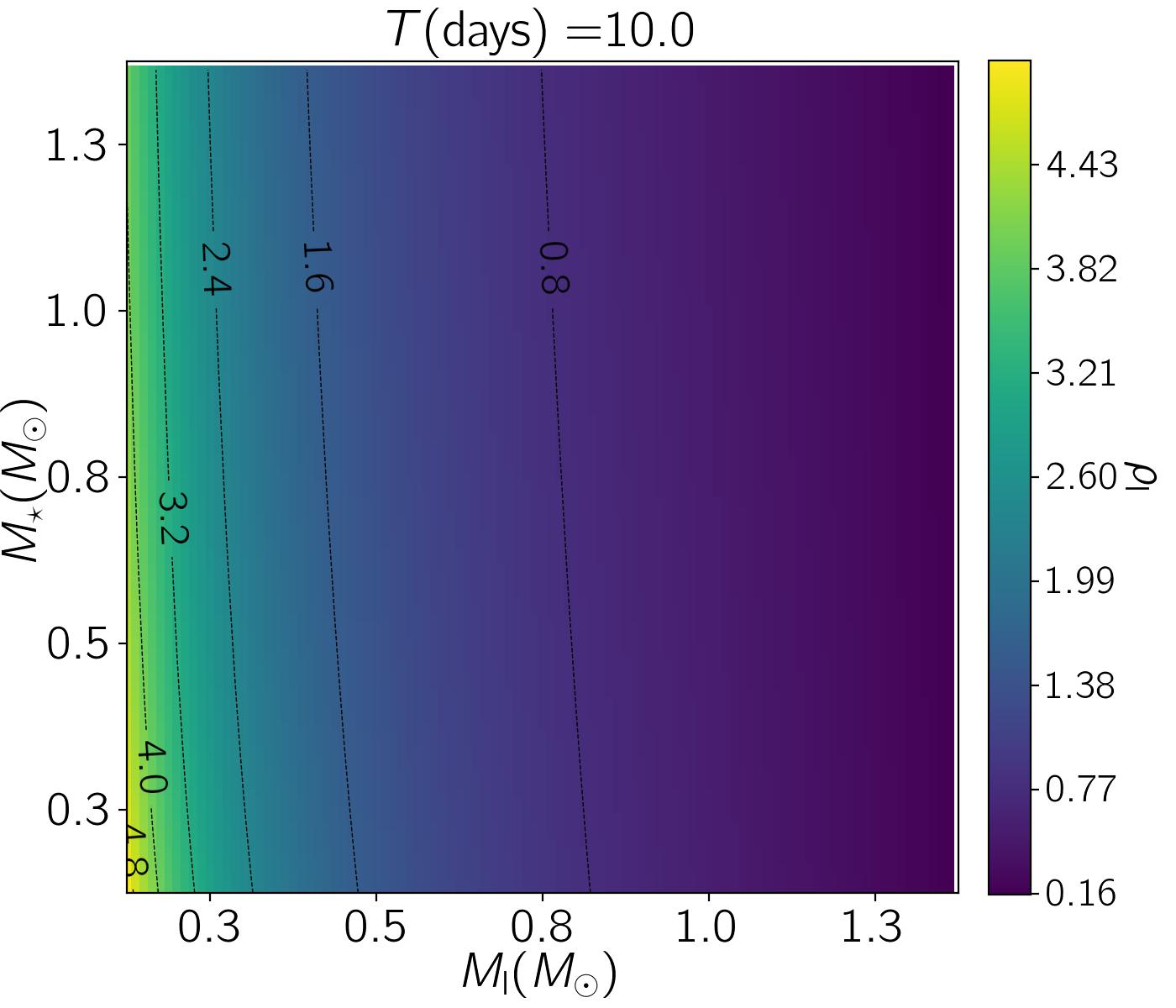}
\caption{The maps of $R_{\rm E}(0.01~R_{\odot})$, $\rho_{\star}$, and $\rho_{\rm{l}}$ over the 2D space $M_{\rm{l}}(M_{\odot})-M_{\star}(M_{\odot})$ due to different DWD systems, respectively. For these maps, two parameters are fixed which are the orbital period $T=10~\rm{days}$, and the system distance from the observer $D_{\rm l}=1~$kpc, and we assume their orbits are circular. To understand the effect of the orbital period on these maps, three animations from the maps of \href{https://iutbox.iut.ac.ir/index.php/s/skQNXEaALCjFsjN}{$R_{\rm E}$}, \href{https://iutbox.iut.ac.ir/index.php/s/i9QRoW7BbR9yfS2}{$\rho_{\star}$}, and \href{https://iutbox.iut.ac.ir/index.php/s/EojfeE3PXB9XkFe}{$\rho_{\rm l}$} versus orbital period are available.}\label{fig1}
\end{figure*}

\section{Lensing/Eclipsing Signals in Edge-on DWD Systems}\label{sec3} 
We explained our method for simulating binary systems including compact objects and stellar companions and making their self-lensing and eclipsing signals by details in \citet{2024sajadianAS,2014MNRASorbit}. Here, for simulate DWD systems and their light curves we use the same method, and review it here briefly. 
To simulate a binary system including two WDs, we consider some parameters which are their masses ($M_{1}$, and $M_{2}$), their surface temperatures ($T_{\rm{eff},~1}$ and $T_{\rm{eff},~2}$), their orbital period $T$, and the orbital eccentricity $\epsilon$. We determine the radii of WDs using the known mass-radius relationship for WDs as given by \citet{1972ApJWDMR}. Also, Kepler's third law indicates the semi-major axis of their orbit $a$.

To project their orbital planes on the sky plane, generally two projection angles are necessary, (i) one to project the minor (or major) axis of the orbital plane on the sky plane $\theta$, and (ii) another one to project the orbital plane normal to the sky plane and on the line of sight toward the observer $i$ which is the so-called inclination angle.  

The self-lensing signals happen when two components of the binary system are collinear (their projected separation on the sky plane is minimum) as seen by the observer. In DWD systems, both components are luminous and compact, so that two self-lensing signals occur during one orbital period, which are not necessarily similar. For each self-lensing signal the foreground WD is the lens object with the mass and radius $M_{\rm l}$, $R_{\rm l}$, and the farther WD is the source object whose mass and radius are specified by $M_{\star}$, $R_{\star}$, respectively.

The self-lensing formalism is similar to the microlensing one, in which we have a characteristic length, the so-called Einstein radius (the radius of the images' ring at the time of the complete alignment) as given by:  
\begin{eqnarray}
R_{\rm E}= \sqrt{\frac{4 G M_{\rm l}}{c^{2}} \frac{D_{\rm ls} D_{l}}{D_{\rm s}}}\simeq 0.01R_{\sun}\sqrt{\frac{M_{\rm l}}{0.6M_{\sun}}\frac{a}{0.1\rm{AU}}},
\end{eqnarray} 
where, $G$ is the Gravitational constant, $c$ is the light's speed, $D_{\rm l}$, and $D_{\rm s}$ are the lens and source distances from the observer, respectively. Also, $D_{\rm ls}=D_{\rm s}-D_{\rm l}$ is the lens-source separation in the line of sight direction. We note that for binary systems $D_{\rm l}\simeq D_{\rm s}$, and $D_{\rm ls}\simeq a$. Accordingly, in DWD binary systems three lengths $R_{\rm l}$, $R_{\rm s}$, and $R_{\rm E}$ have similar orders of magnitude. 

In DWD systems, the magnification factor suffers from two effects finite-source size and finite-lens size. The finite-source size is evaluated by $\rho_{\star}$ which is the source radius projected on the lens plane and normalized to the Einstein radius as given by:
\begin{eqnarray}
\rho_{\star}=\frac{R_{\star}D_{\rm l}}{D_{\rm s} R_{\rm E}}\simeq 1 \frac{0.01R_{\sun}}{R_{\star}}\frac{R_{\rm E}}{0.01R_{\sun}}
\end{eqnarray} 
\noindent The magnification factor by considering the finite-source size which was first calculated by \citet{1994ApJwittmao} as given by: 
\begin{eqnarray}
A(\rho_{\star},~u)&=&\frac{1}{\pi}\Big[-\frac{u-\rho_{\star}}{\rho_{\star}^{2}}\frac{8+u^{2}-\rho_{\star}^{2}}{\sqrt{4+(u-\rho_{\star})^{2}}}~F\big(\frac{\pi}{2},~k\big)\nonumber\\
&+&\frac{u+\rho_{\star}}{\rho_{\star}^{2}}~\sqrt{4+(u-\rho_{\star})^{2}}~E\big(\frac{\pi}{2},~k\big)\\
&+&\frac{4(u-\rho_{\star})^{2}}{\rho_{\star}^{2}(u+\rho_{\star})}~\frac{1+\rho_{\star}^{2}}{\sqrt{4+(u-\rho_{\star})^{2}}}~\Pi\big(\frac{\pi}{2},~n,~k\big)\Big]\nonumber,
\end{eqnarray}
where, $u$ is the lens-source separation projected on the lens plane and normalized to the Einstein radius, $n=4 u \rho_{\star}/(u+\rho_{\star})^{2}$, and $k=\sqrt{4 n/\big(4+(u-\rho_{\star})^{2}\big)}$. Also, the functions $F$, $E$, and $\Pi$ are the first, second, and third types of the elliptical integral, respectively. For extremely large source sizes ($\rho_{\star}\gg 1$) and when the lens is crossing the source disk the magnification factor is estimated by $A\simeq 1+ 2/\rho_{\star}^{2}$ and does not depend on $u$ \citep{1973Maeder,1996Gould,2003ApJAgol,2023sajadiandegeneracy}.Generally, enhancing finite-source size decreases the magnification factor and generates a more-flattened light curve. We calculate the magnification factor during self-lensing signals and by considering the finite-source size and a limb-darkened profile for the source surface brightness, $A(\rho_{\star},~u,~\Gamma)$, using the public code \texttt{VBBinaryLensing} \citep{2010MNRASBozza,2018MNRABozza}. Here, $\Gamma$ is the linear limb-darkening coefficient.

The finite-lens size is evaluated by the normalized lens radius which is: 
\begin{eqnarray}
\rho_{\rm l}=\frac{R_{\rm l}}{R_{\rm E}}\simeq 1\frac{0.01R_{\sun}}{R_{\rm l}}\frac{R_{\rm E}}{0.01R_{\sun}}.
\end{eqnarray} 
The finite-lens effect reduces the magnification factor because the lens's disk blocks some parts of the images's area. We show the occultation due to finite-lens effect by $\mathcal{O}$ which is the ratio of the images' area which is covered by the lens's disk to the source area (when both of them are projected on the lens plane) and calculate it numerically using the inverse-ray-shooting (IRS) method \citep[see, e.g.,  ][]{1998LRRWambsganss,2010MNRAsajadian}.
This factor is given by:
\begin{eqnarray}
\mathcal{O}=\frac{1}{\pi R_{\star,\rm{p}}^{2}(1-\Gamma/3 )}~\int_{I}dx~dy~f(x,~y,~\Gamma)~\Theta(R),
\end{eqnarray}	
where, the integration is done over the images' area ($I$) projected on the lens plane and specified with $(x,~y)$, $R=\sqrt{x^{2}+y^{2}}$ is the distance of a given point over the images' area from the lens position (which is at the center of the coordinate system), $\Theta$ is a step function which is one if $R\leq R_{\rm{l}}$, and it is zero otherwise. $f(x,~y,~\Gamma)$ is the surface brightness over the images's disk at the position $(x,~y)$. Also, $R_{\star, \rm{p}}=R_{\star} D_{\rm l}/D_{\rm s}$ is the source radius projected on the lens plane.

\noindent The fraction of the images' area covered by the lens disk is given by $f_{\mathcal{O}}=\mathcal{O}\big/A$. If $f_{\mathcal{O}}=1$ the images are completely covered by the lens and a complete eclipse happens. The overall normalized fluxes while lensing signals by the first and second WD are (respectively):
\begin{eqnarray}
\mathcal{A}_{1}&=& \frac{(A-\mathcal{O})+\mathcal{F}}{1+\mathcal{F}},\nonumber\\
\mathcal{A}_{2}&=& \frac{1+(A-\mathcal{O})\mathcal{F}}{1+\mathcal{F}},
\label{magni}
\end{eqnarray}  
where $\mathcal{F}$ is the ratio of the first WD's flux to the second WD's flux.

In Figure \ref{fig1}, we show the maps of $R_{\rm E}(0.01R_{\sun})$, $\rho_{\star}$, and $\rho_{\rm l}$ over the 2D space $M_{\rm l}(M_{\odot})-M_{\star}(M_{\odot})$ due to different DWD systems, in three panels. The ranges of lens and source masses are $[0.17,~1.4] M_{\odot}$ with $100$ grids over each axis. We fix their orbital periods to $10$ days, the system distance from the observer to one kpc, and the orbital eccentricity to zero for these maps. To find the effect of the orbital period on these maps, we make them for a wide range of the orbital period $T\in[1,~50]$ days, and three animations from the resulting maps versus $T$ are available. 

According to these figures, $\rho_{\rm l}$ values depend strongly on the lens object as well as the orbital period, and does not significantly depend on source stars. For massive WDs, e.g., $M_{\rm l}\gtrsim1.3 M_{\odot}$, $\rho_{\rm l}\sim 0.3-0.4$ even for a wide range of the orbital period, so that we do not expect significant eclipsing effects. When lens and source objects are low-mass ones with $M\lesssim 0.3 M_{\odot}$, $\rho_{\star}$, and $\rho_{\rm l}$ are as large as $\gtrsim 3$, so no self-lensing signal occurs, and a complete eclipse for short orbital periods will happen. Reversely, DWD systems including two massive WDs with masses higher than a solar mass have $\rho_{\rm l}\lesssim 1$ and $\rho_{\star}\lesssim1$ which offer self-lensing signals without eclipsing.  
\begin{figure*}
\centering
\includegraphics[width=0.329\textwidth]{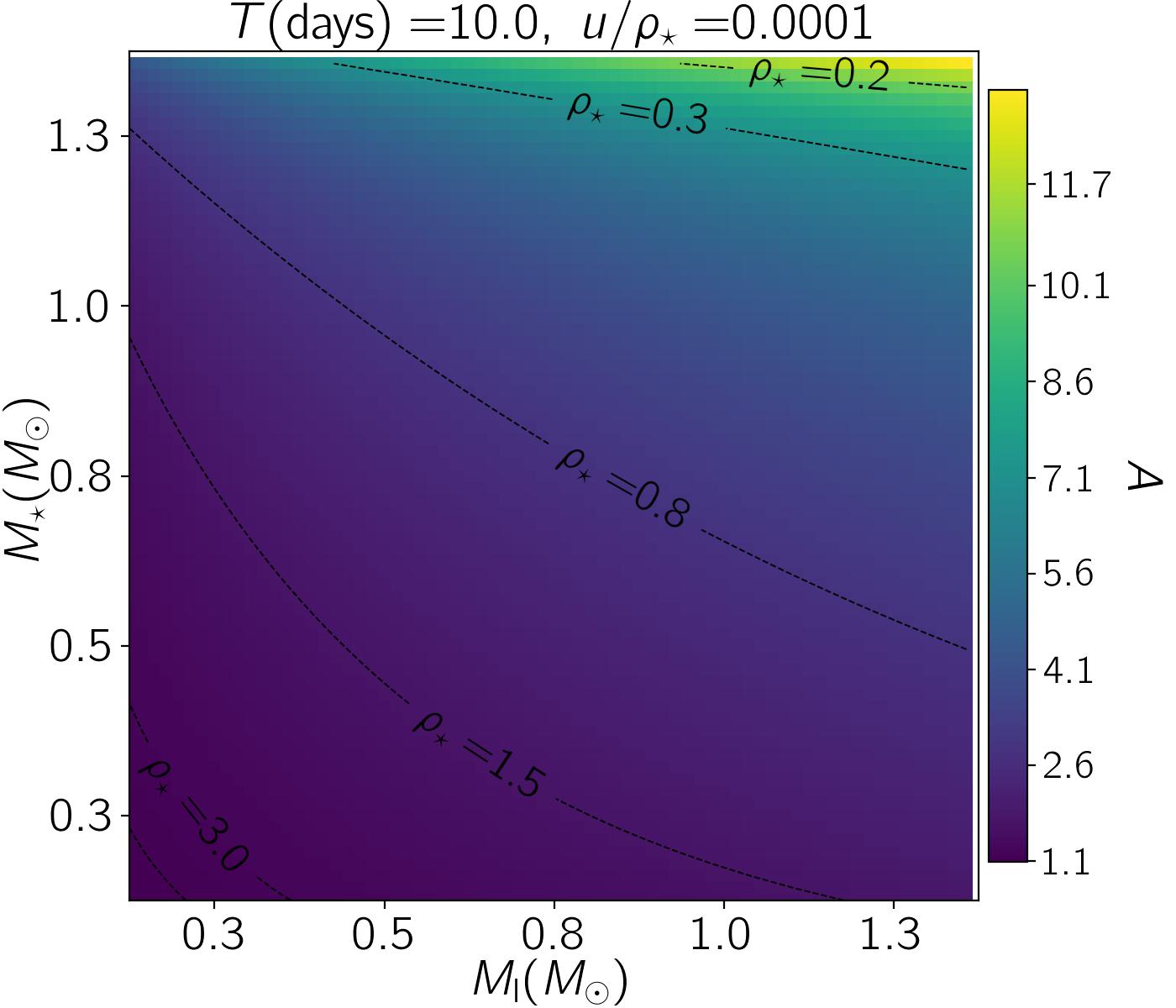}
\includegraphics[width=0.329\textwidth]{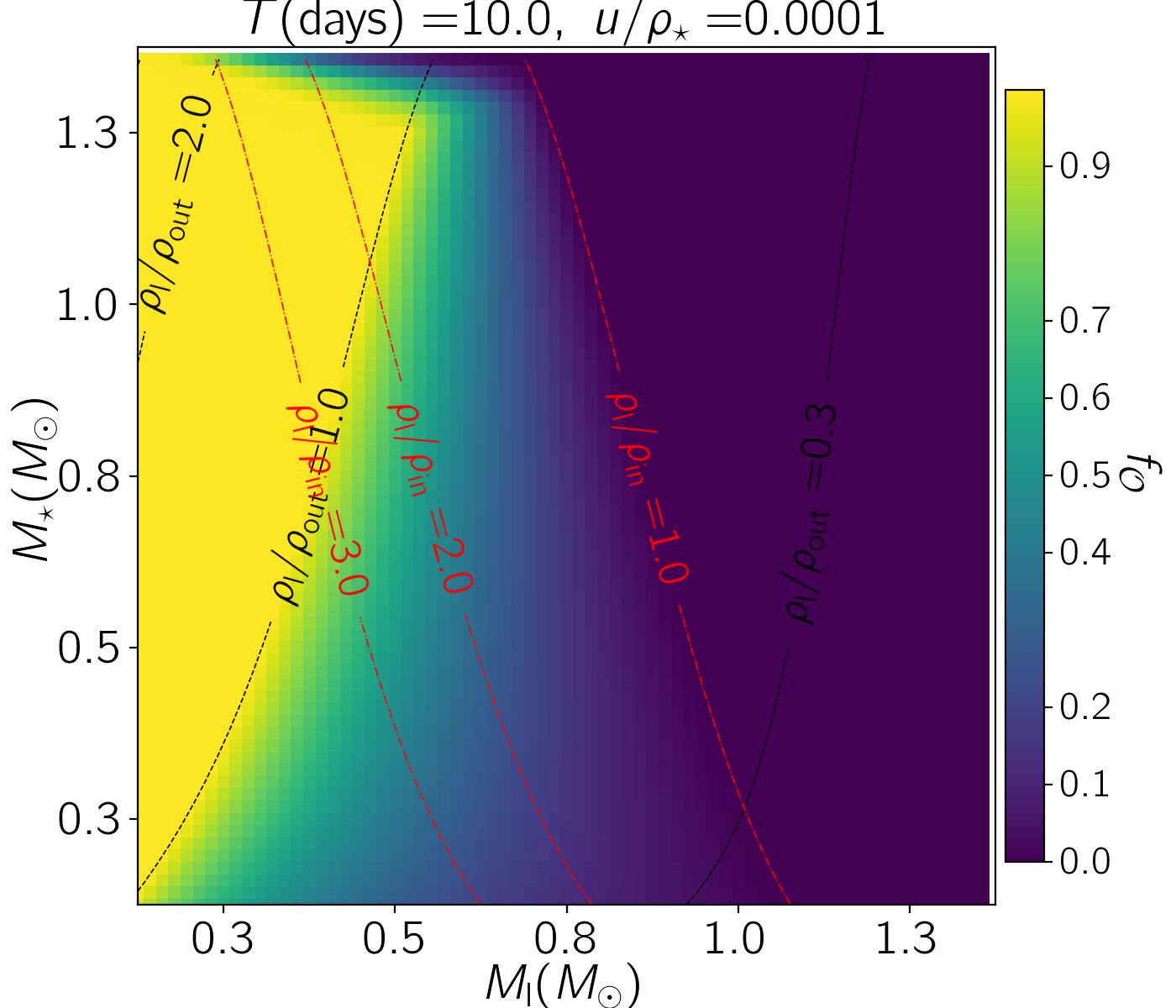}
\includegraphics[width=0.329\textwidth]{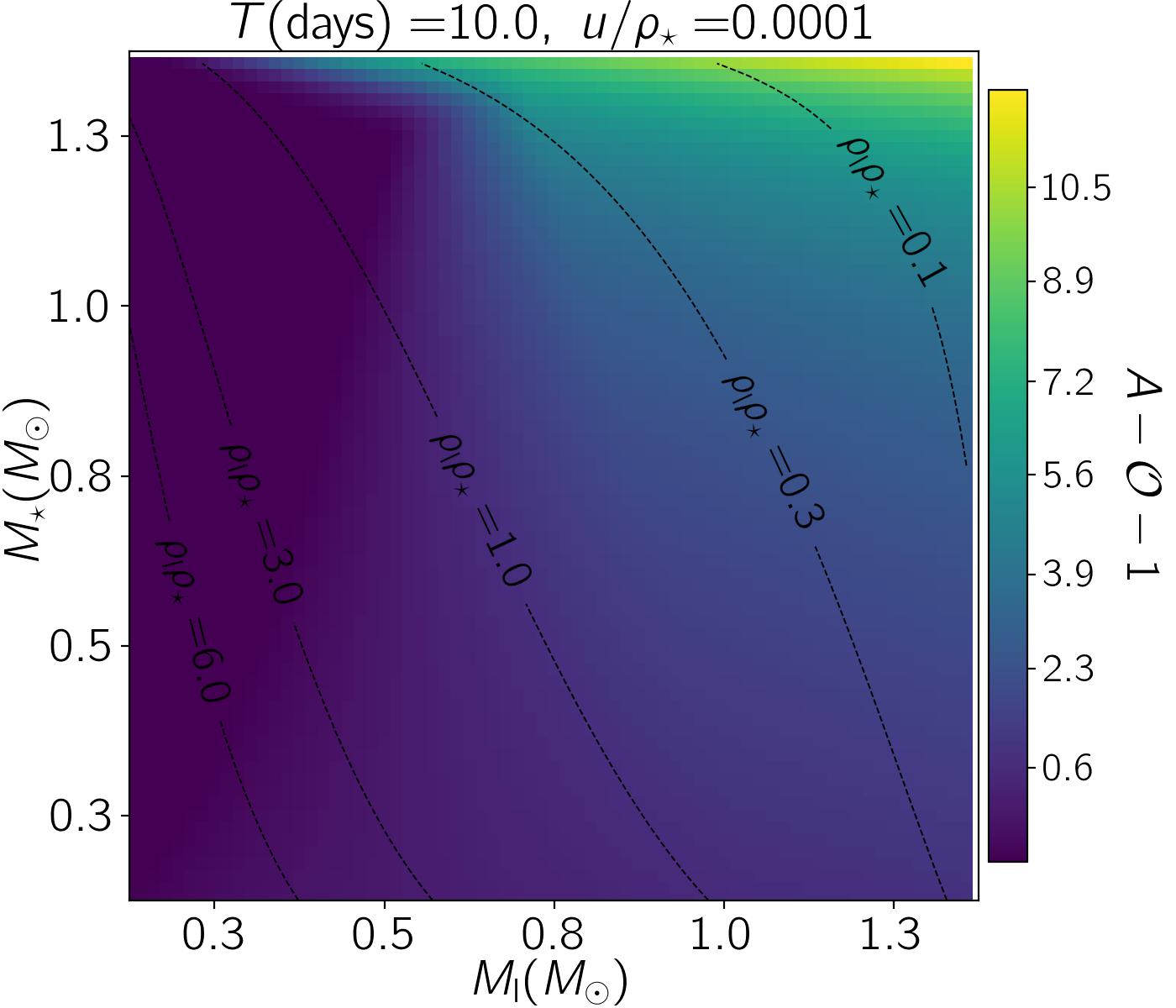}
\includegraphics[width=0.329\textwidth]{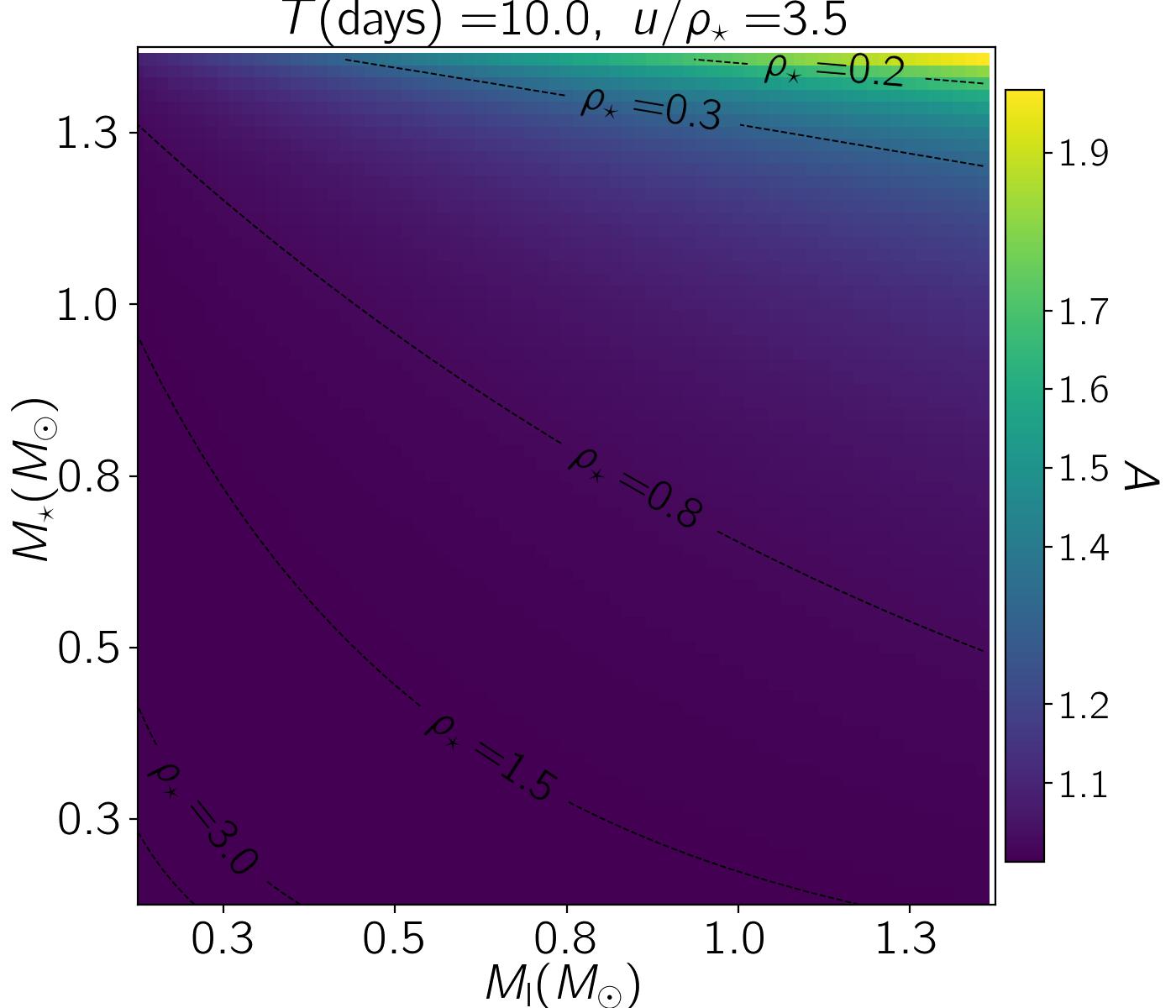}
\includegraphics[width=0.329\textwidth]{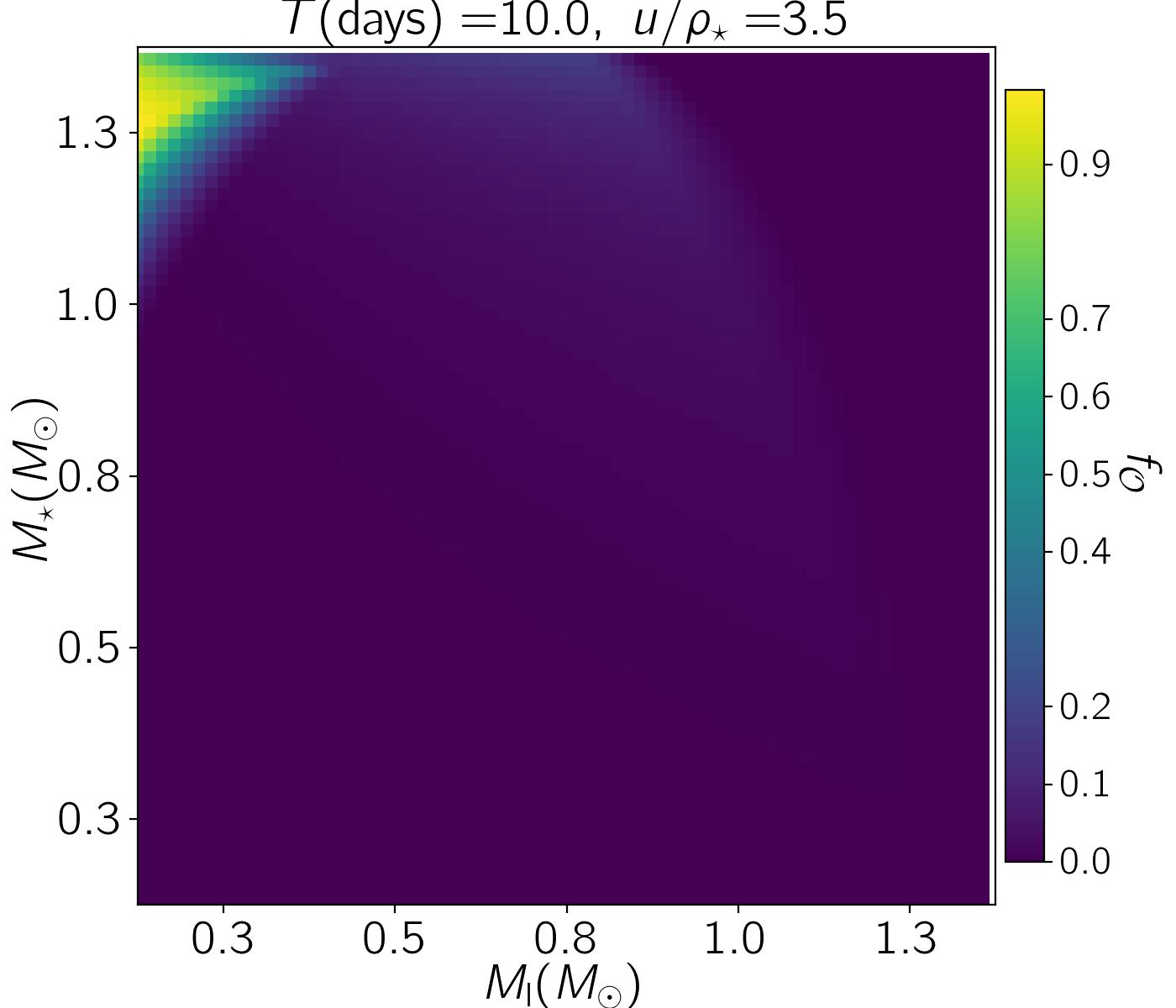}
\includegraphics[width=0.329\textwidth]{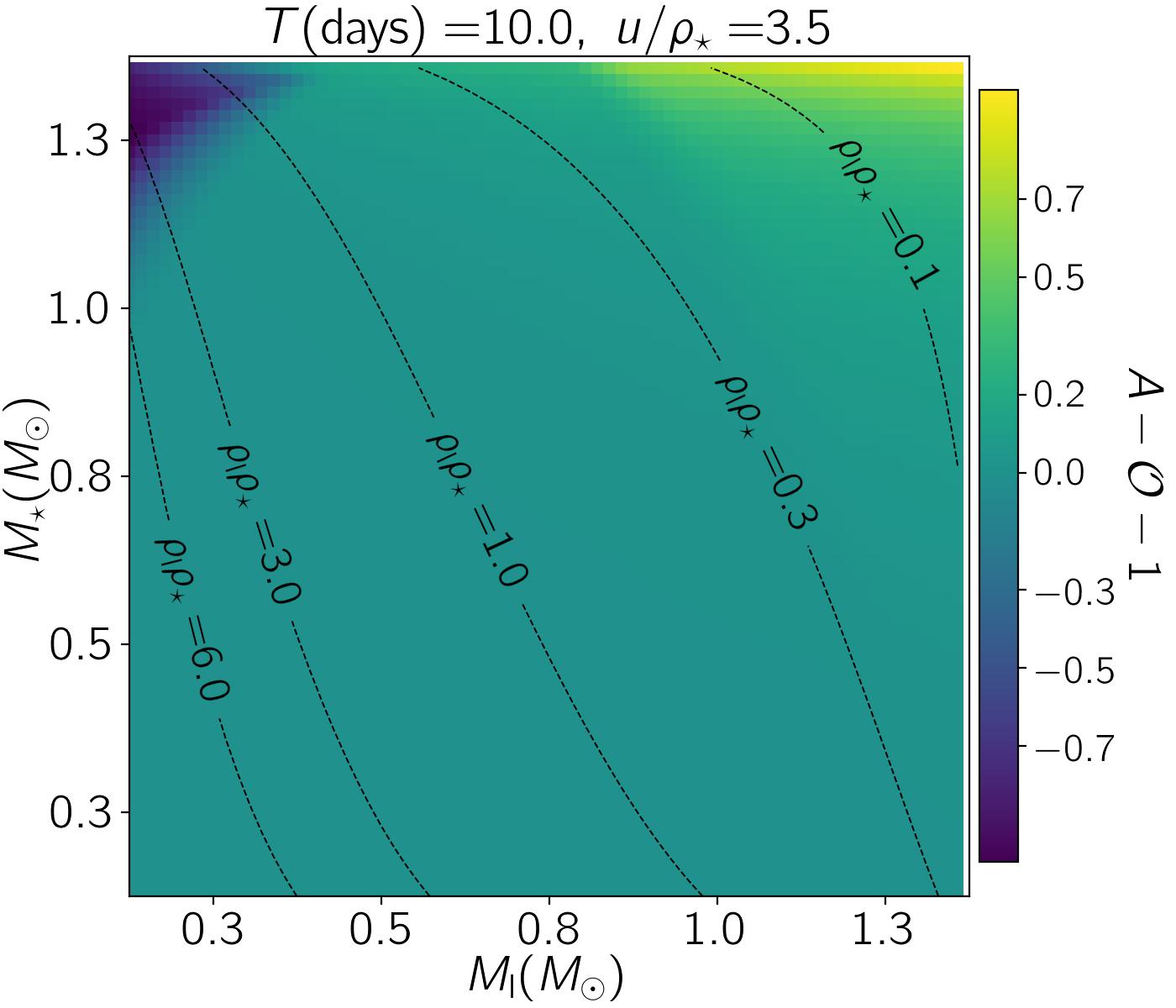}
\caption{The maps of the magnification factor $A$, the fraction of the images' area covered by the lens object $f_{\mathcal{O}}$, and $\delta=A-\mathcal{O}-1$ which is the relative residual in the normalized flux due to both self-lensing and occultation signals over the 2D space $M_{\rm{l}}(M_{\odot})-M_{\star}(M_{\odot})$. Over these maps the contour lines of $\rho_{\star}$ (left panel), $\rho_{\rm l}/\rho_{\rm in}$ and $\rho_{\rm l}/\rho_{\rm{out}}$ (middle panel), and $\rho_{\rm l}\rho_{\star}$ (right panel) are shown. For these maps some parameters are fixed which are $T=10$ days, $D_{\rm l}=1$ kpc, $\Gamma=0.45$, and the lens-source separation normalized to the Einstein radius is $u=10^{-4}\rho_{\star}$ (three top panels) and $u=3.5\rho_{\star}$ (three bottom panels). To study the effect of the orbital period on these maps, for $T\in [1,~50]$ days we generate these maps and three animations from the maps of \href{https://iutbox.iut.ac.ir/index.php/s/F8oZA93d8C784Xi}{$A$}, \href{https://iutbox.iut.ac.ir/index.php/s/pqt7BxLAyic3jNE}{$f_{\mathcal{O}}$}, and \href{https://iutbox.iut.ac.ir/index.php/s/FzDtjNpeW9pZmCG}{$\delta$} are available.}\label{fig2}
\end{figure*}

In Figure \ref{fig2}, we show the maps of the magnification factor $A$, the fraction of the images' area covered by the lens object $f_{\mathcal{O}}$, and $\delta=A-\mathcal{O}-1$ which is the relative residual (the ratio of the residual area between the images and the source star to the source's area) over the 2D space $M_{\rm l} (M_{\odot})-M_{\star}(M_{\odot})$ and by considering two values for the lens-source separation normalized to the Einstein radius $u=10^{-4}\rho_{\star}$ (top panels) and $u=3.5\rho_{\star}$ (bottom panels). Here some parameters are fixed which are $T=10$ days, $\Gamma=0.45$, $D_{\rm l}=1$ kpc. However, for other orbital periods in the range $T\in [1,~50]$ days we make these maps and three animations from them are available.

\noindent Here, the relative residual will be positive, i.e., $\delta>0$, when the normalized flux of the source star (the ratio of the images' flux received by the observer to the stellar flux at the baseline) is greater than one. If the normalized flux of the source star is less than one, then $\delta<0$. The minimum value of $\delta$ is $-1$ which occurs when the images' disk is completely covered by the lens's area, i.e., $A=\mathcal{O}$. Also, $\delta=0$ occurs when there are no net lensing/eclipsing signals for the source star. In addition, over the map of magnification factor the contour lines of $\rho_{\star}$ are plotted with dashed black curves. The magnification factor in DWD systems due to massive WDs is as high as $10$ or even  more, because for these systems $\rho_{\star}$ is considerably small and $\sim 0.1$.

When the lens is crossing the source disk ($u_{0}<\rho_{\star}$), the fraction of the images' disk which is covered by the lens $f_{\mathcal{O}}$ can be evaluated by comparing the inner and outer radii of the images' ring ($R_{\rm in}$ and $R_{\rm out}$) and the lens radius. Here, $u_{0}$ is the lens impact parameter which is the minimum lens-source separation projected on the lens plane and normalized to the Einstein radius. The inner and outer radii of the images' ring normalized to the Einstein radius (when $\rho_{\star}$ is not small) are given by $\rho_{\rm{in}}\sim \big(\sqrt{\rho_{\star}^{2}+4}-\rho_{\star}\big)/2$ and $\rho_{\rm{out}}\sim \big(\sqrt{\rho_{\star}^2+4}+\rho_{\star}\big)/2$, respectively. If $\rho_{\rm l}>\rho_{\rm{out}}$ a complete eclipse occurs, and when $\rho_{\rm l}<\rho_{\rm in}$ no eclipse happens. Between these two limits (i.e., $\rho_{\rm in }\leq\rho_{\rm l}\leq\rho_{\rm{out}}$) a partial eclipse occurs.

Hence, over the map of $f_{\mathcal{O}}$ (the top middle map in Figure \ref{fig2}) the contour lines of $\rho_{\rm l}/\rho_{\rm{out}}$ and $\rho_{\rm l}/\rho_{\rm{in}}$ are plotted with dashed black and dot-dashed red curves, respectively. For $\rho_{\rm l}>\rho_{\rm out}$, complete eclipse happens, i.e., $f_{\mathcal{O}}=1$ (yellow parts). Accordingly, in transit events when $M_{\rm l}\lesssim 0.3 M_{\odot}$ complete eclipses occur, and for $M_{\rm l}\gtrsim1.0 M_{\odot}$ no eclipse happens. 

The magnification factor and eclipsing signals depend on $\rho_{\star}$ and $\rho_{\rm l}$, respectively, so that when both of them are reduced self-lensing and eclipse are enhanced and decreased, respectively. Hence, on the map of the relative residual $\delta$ we show the contour lines of $\rho_{\rm l}\times\rho_{\star}$ whose amount reduces by the orbital period and both masses of the source and lens WDs. Hence, DWDs including two massive WDs in wide orbits have highest magnification factors, and DWDs including one ELM WD (as the lens object) and one massive WD (as the source star) in close orbits have deepest or complete eclipsing signals ($-1\leq\delta\leq0$).    

In three bottom panels of Figure \ref{fig2}, we show similar maps for a larger lens-source separation (normalized to the Einstein radius) $u=3.5\rho_{\star}$ in which the lens does not cross the source's disk. The magnification factor can reach $2$ when both the lens and source WDs are massive. The complete eclipse happens only when an ELM WD is the lens object and orbiting a very massive WD (the source star). We do not plot the contour lines $\rho_{\rm l}/\rho_{\rm in}$ and $\rho_{\rm l}/\rho_{\rm out}$ over the middle maps, because for this $u$ value the ring of images is not generated. In the next section, we simulate several light curves due to different edge-on DWD systems and discuss on their properties. 

\begin{figure*}
\centering
\subfigure[]{\includegraphics[width=0.49\textwidth]{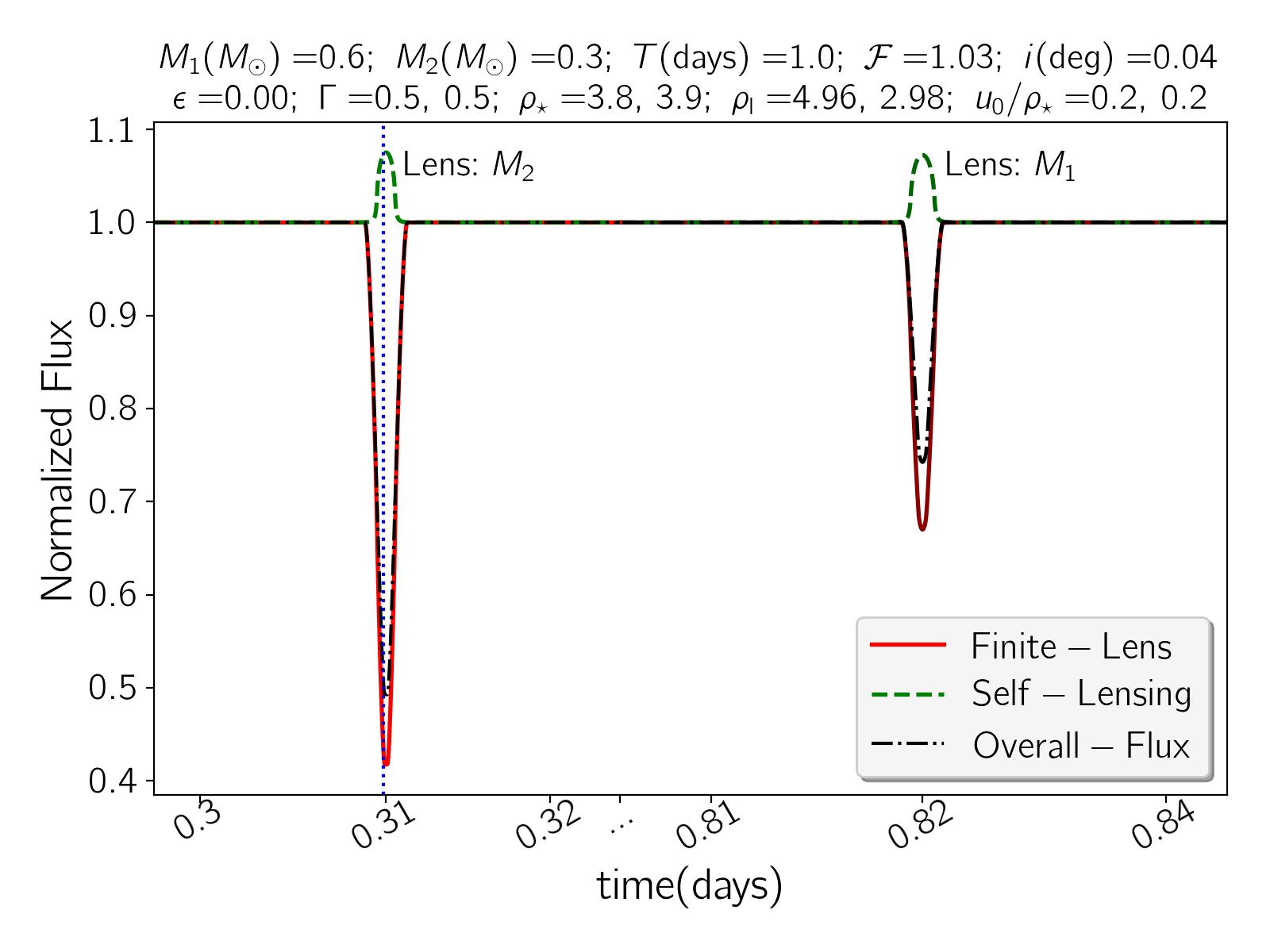}\label{lca}}
\subfigure[]{\includegraphics[width=0.49\textwidth]{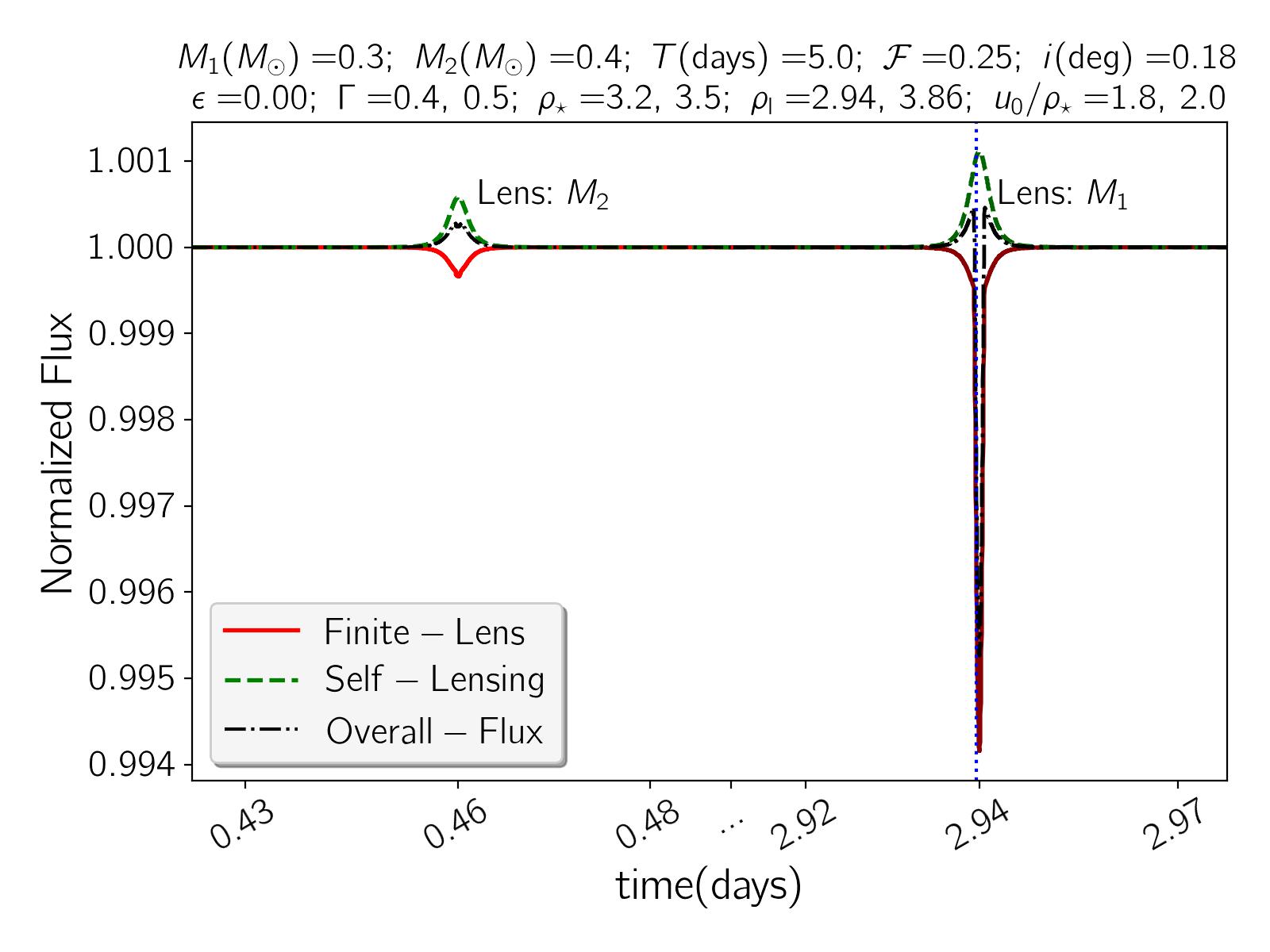}\label{lcb}}\\
\subfigure[]{\includegraphics[width=0.49\textwidth]{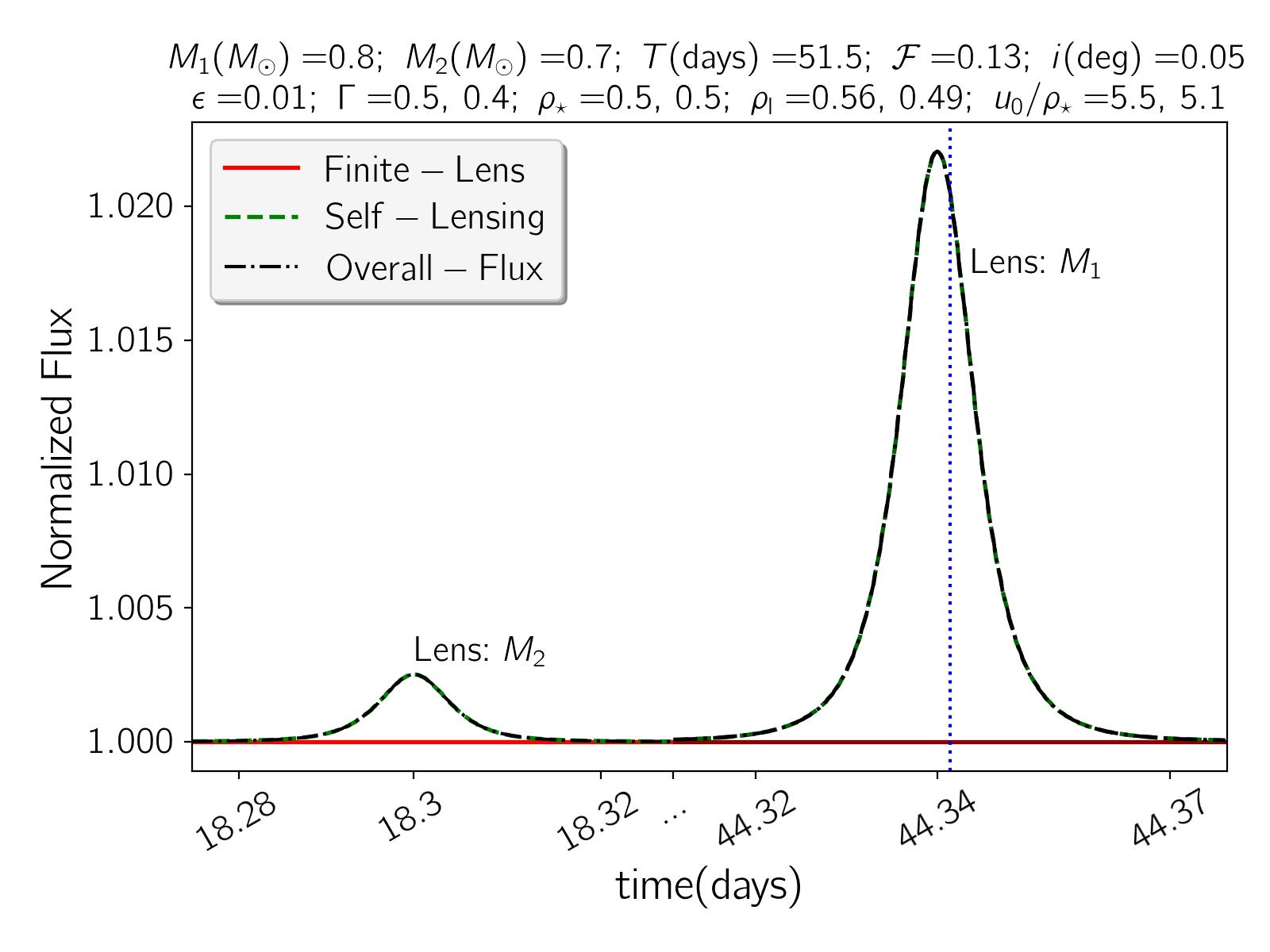}\label{lcc}}
\subfigure[]{\includegraphics[width=0.49\textwidth]{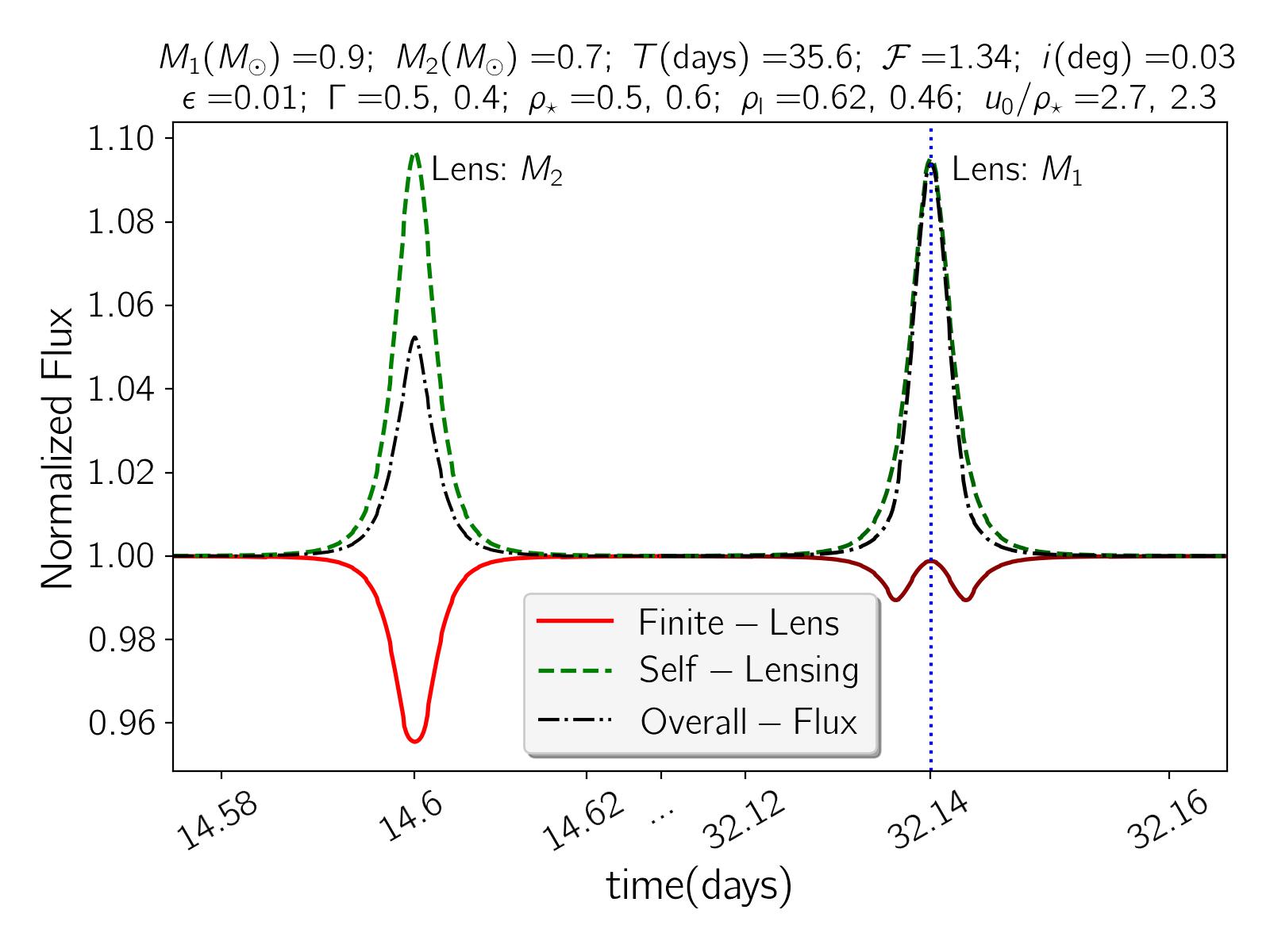}\label{lcd}}\\
\subfigure[]{\includegraphics[width=0.49\textwidth]{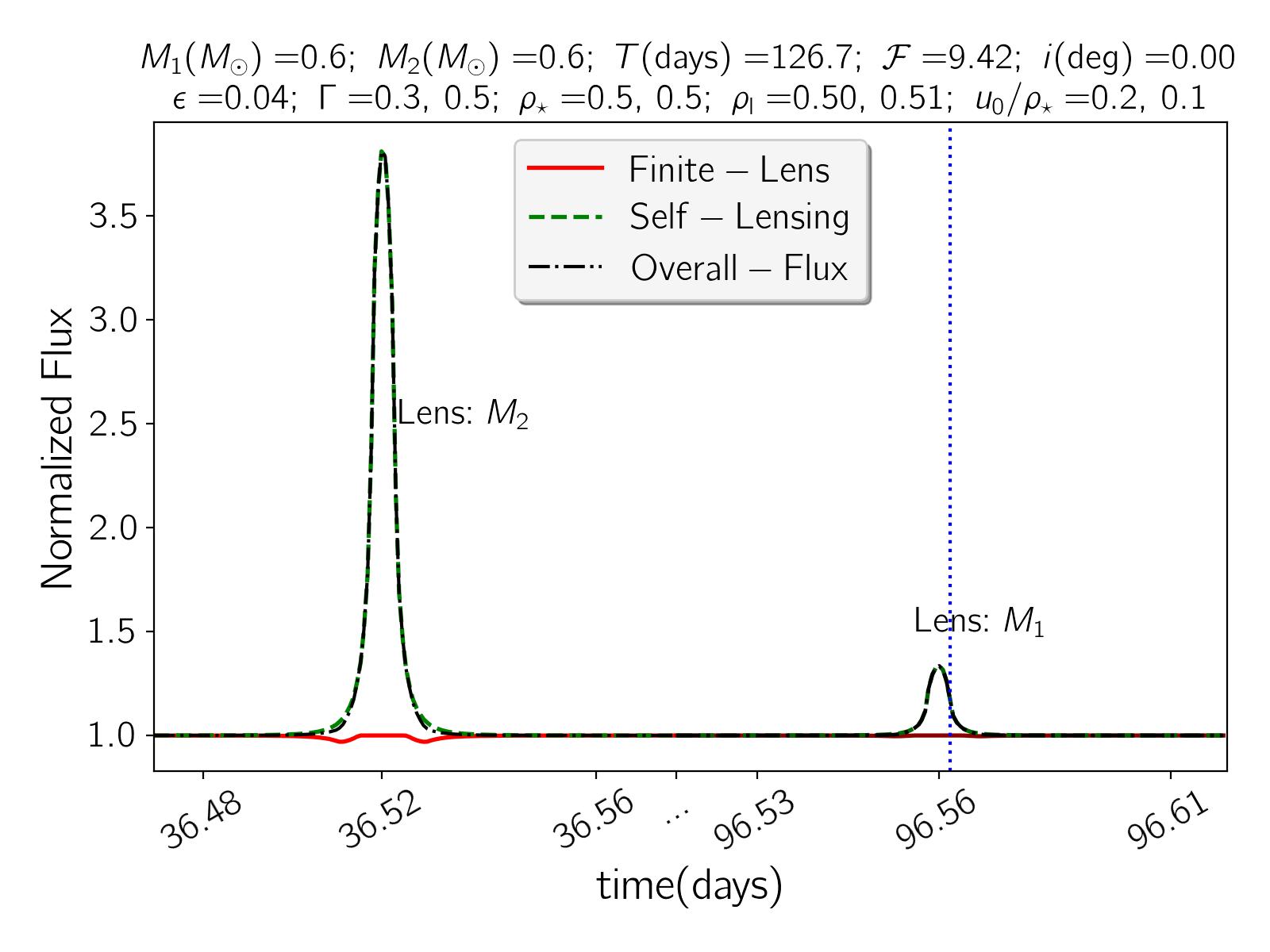}\label{lce}}
\subfigure[]{\includegraphics[width=0.49\textwidth]{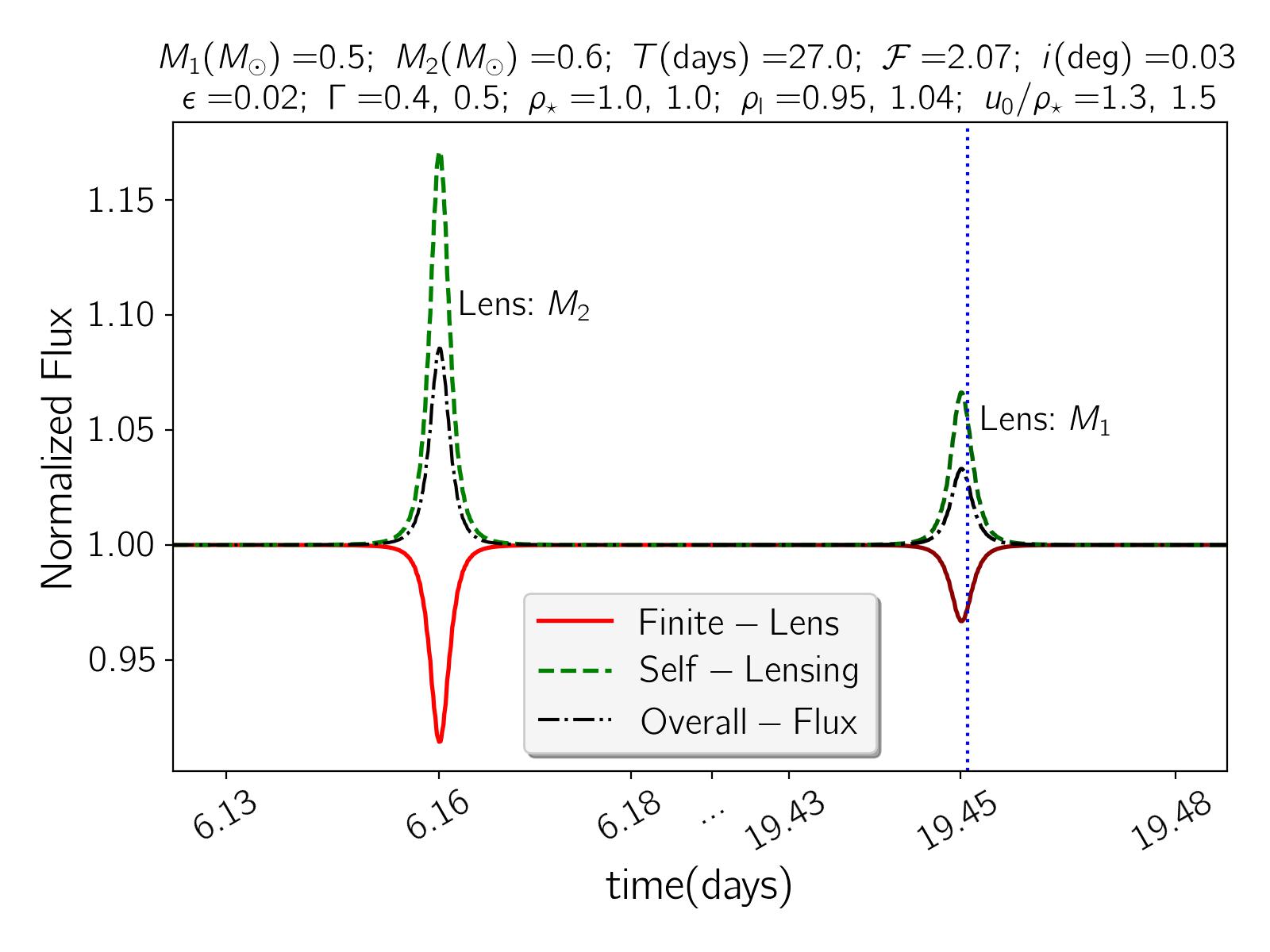}\label{lcf}}
\caption{Six examples of light curves due to different edge-on DWD systems are represented in different panels. In each panel, three curves are shown which are the normalized fluxes by considering self-lensing (dashed green), eclipse by the lens disk (solid red) and both of them (dot-dashed black) versus time. At the top of each panel, some key parameters are mentioned. For $\Gamma$, $\rho_{\star}$, $\rho_{\rm l}$, and $u_{0}/\rho_{\star}$ two values are given which are due to two signals, from left to right respectively. The lensing/eclipsing signals due to the lens object with the mass $M_{1}$ are specified with dark green and dark red colors and labeled with it. The horizontal axes show two zoomed time intervals around lensing/eclipsing signals. The dotted blue lines determine the times at which their corresponding images' surface brightness are shown in Figure \ref{fig4}.}\label{fig3}
\end{figure*}

\section{Light Curves of Edge-on DWD systems}\label{sec4}

\begin{figure*}
\centering
\subfigure[]{\includegraphics[width=0.49\textwidth]{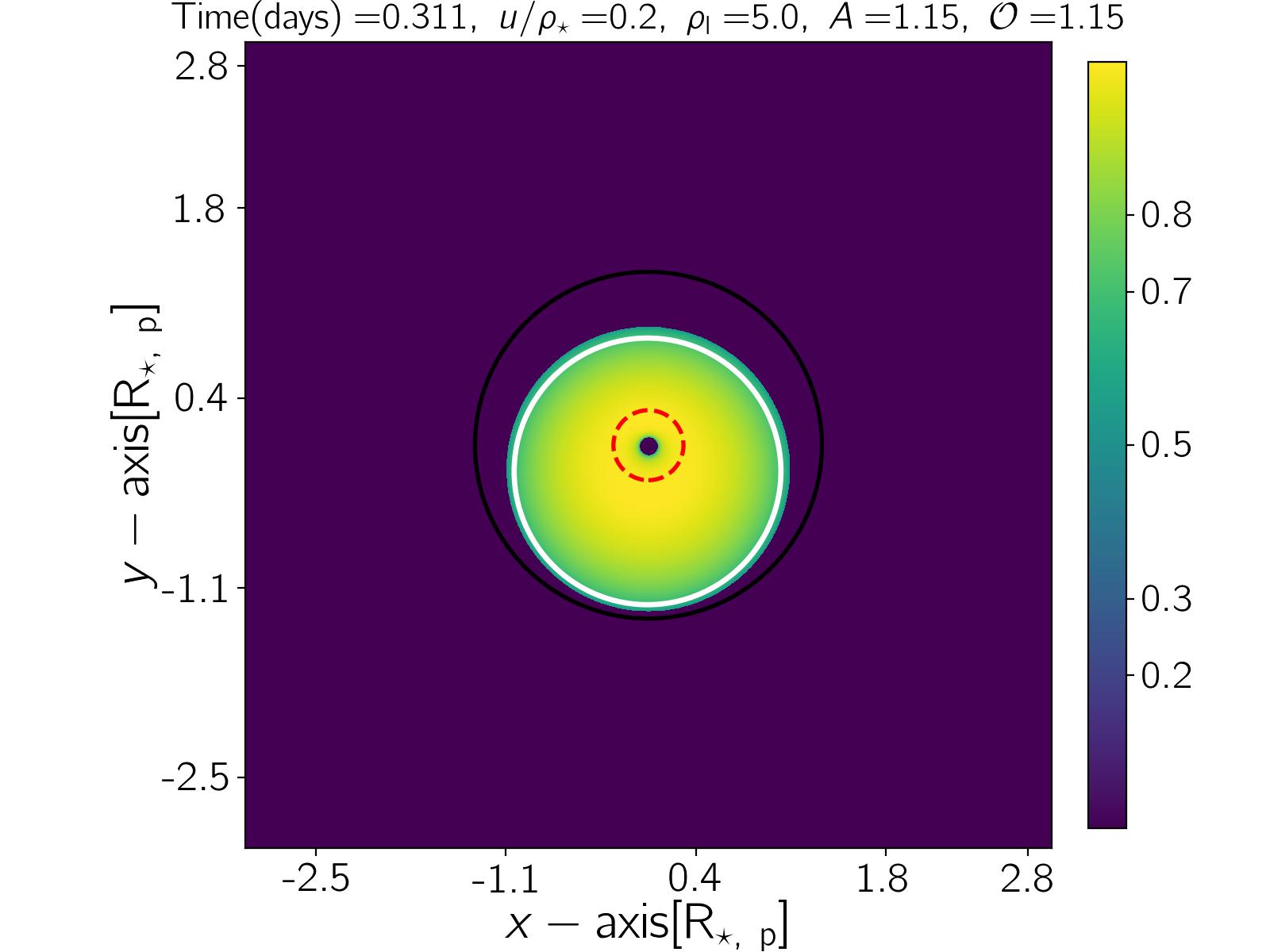}\label{mca}}
\subfigure[]{\includegraphics[width=0.49\textwidth]{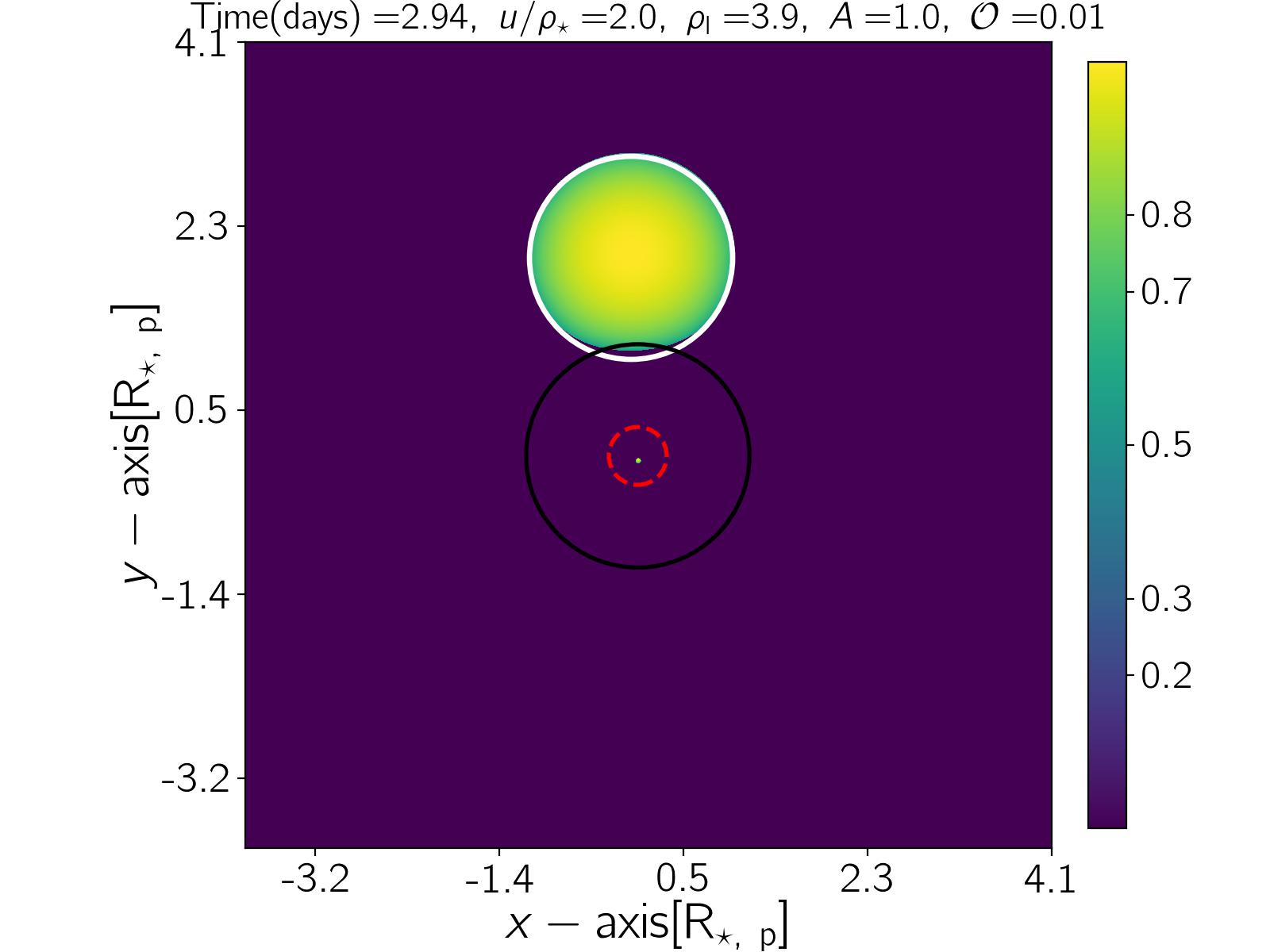}\label{mcb}}\\
\subfigure[]{\includegraphics[width=0.49\textwidth]{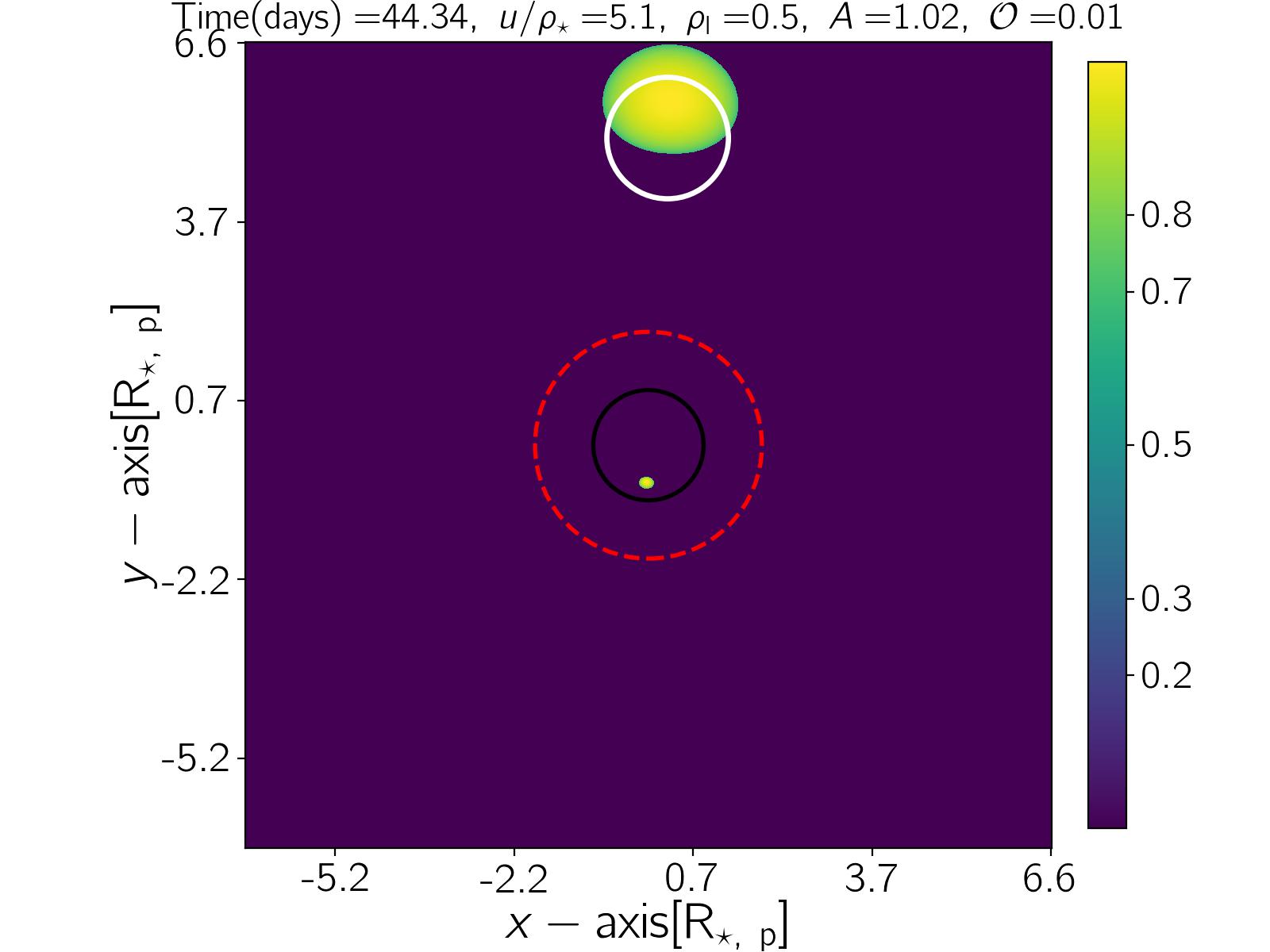}\label{mcc}}
\subfigure[]{\includegraphics[width=0.49\textwidth]{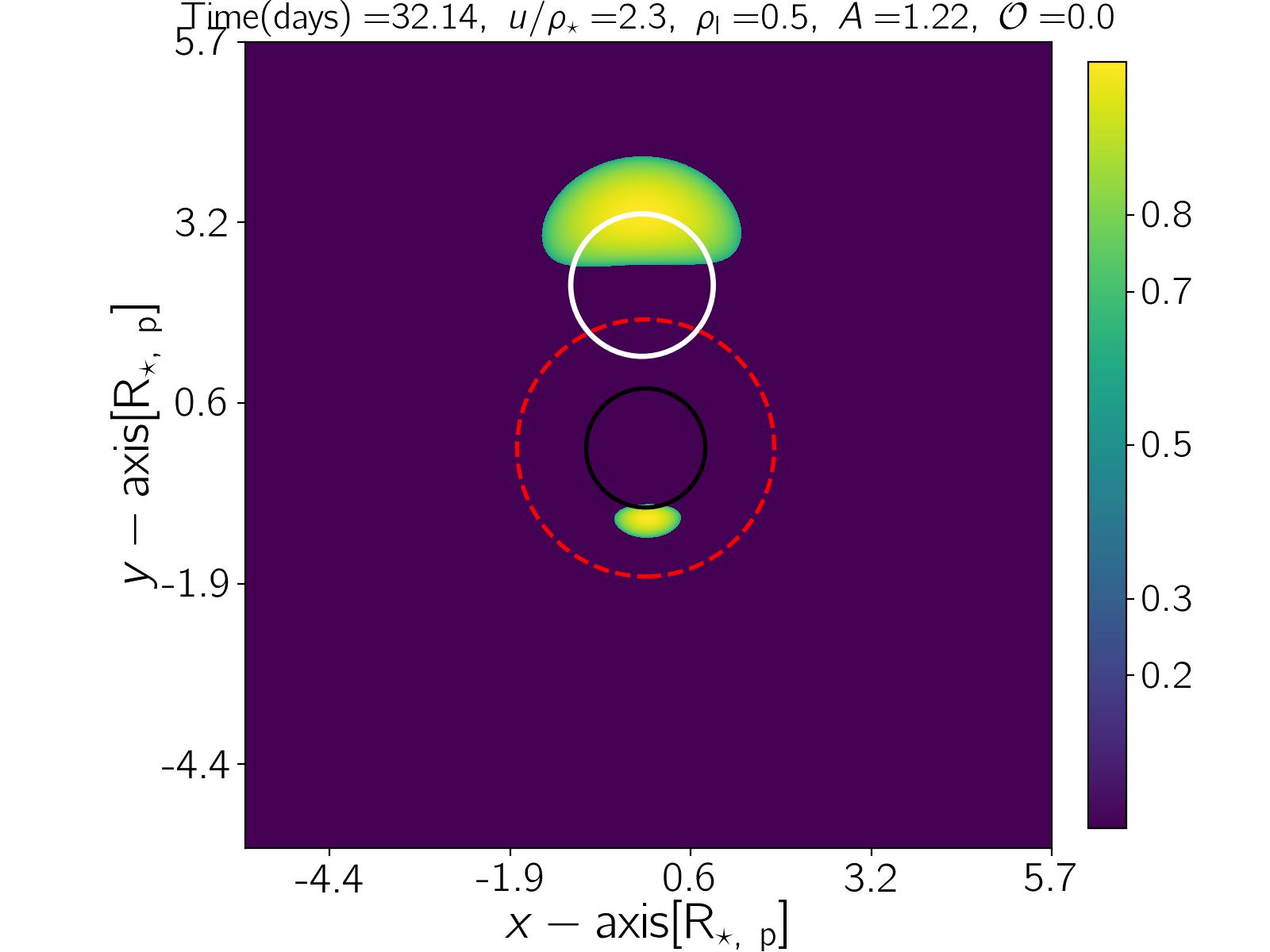}\label{mcd}}\\
\subfigure[]{\includegraphics[width=0.49\textwidth]{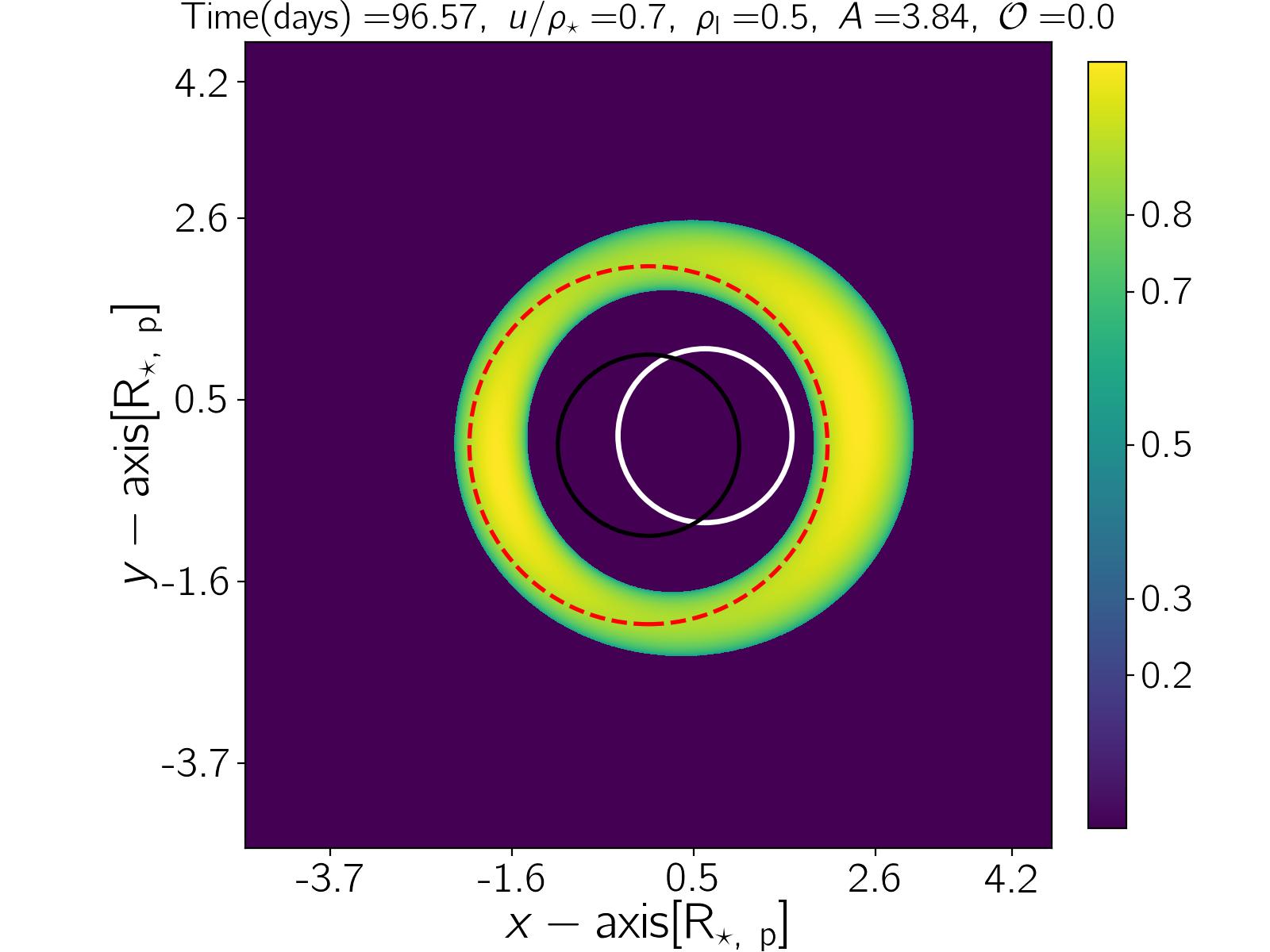}\label{mce}}
\subfigure[]{\includegraphics[width=0.49\textwidth]{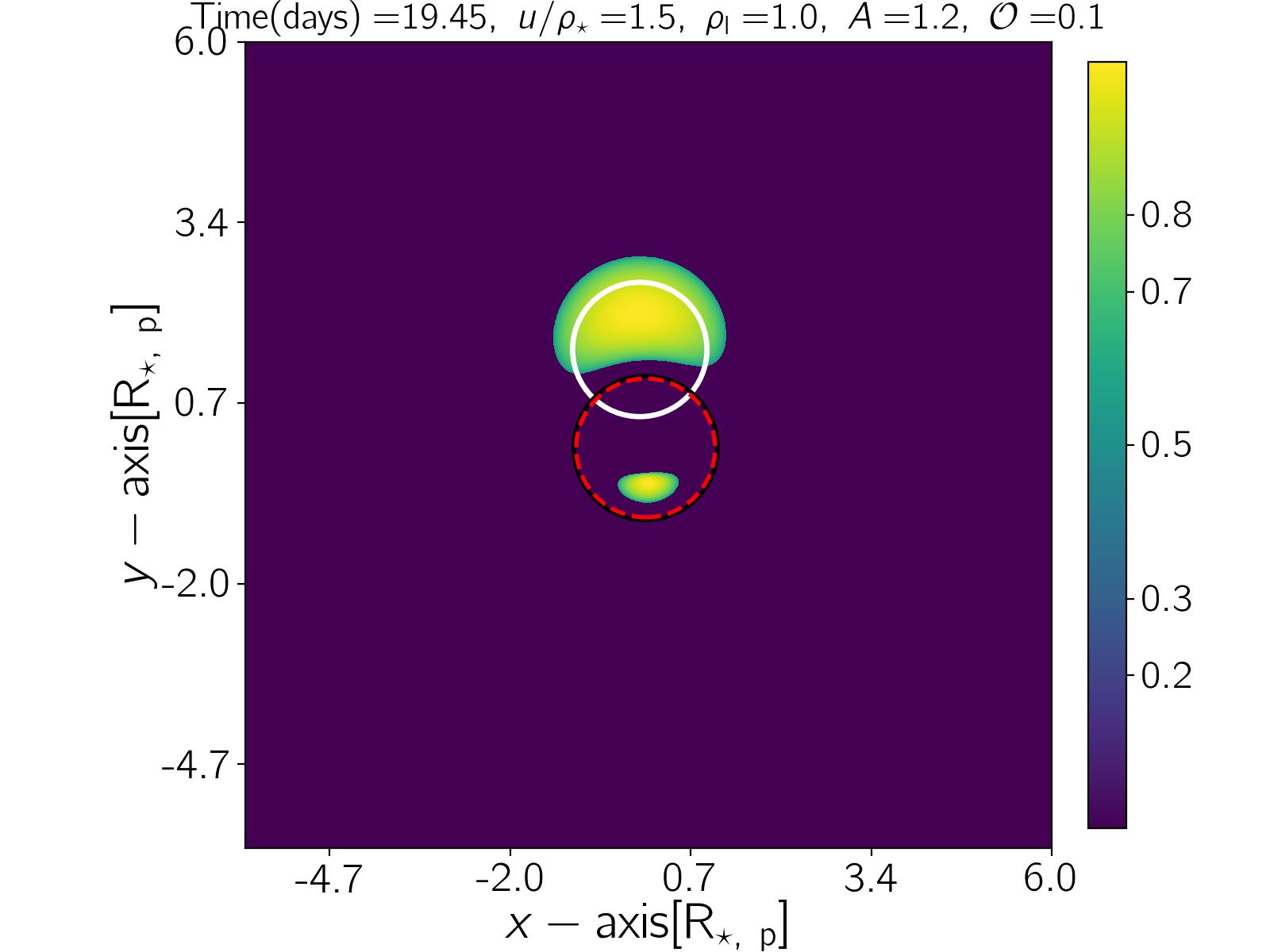}\label{mcf}}
\caption{Each panel shows the lens plane for each light curve in Figure \ref{fig3} at the time specified with a dotted vertical line on that light curve. In each panel, the map determines the lensing-induced images' surface brightness. The white, black and dashed red circles specify the source star' edge, the lens's edge and the Einstein ring, respectively. For each panel some relevant parameters are mentioned on its top. The axes are normalized to the source radius projected on the lens plane $R_{\star,~\rm{p}}=R_{\star}D_{\rm l}/D_{\rm s}$. For each light curve, we make more maps versus time, and six animations from these maps related to the light curves in Figures \href{https://iutbox.iut.ac.ir/index.php/s/LLLW2Xbrs66DWsg}{3(a)}, \href{https://iutbox.iut.ac.ir/index.php/s/nnRrakCyZa9bpFc}{3(b)}, 
\href{https://iutbox.iut.ac.ir/index.php/s/WoGPsSPrGRrwNRj}{3(c)}, \href{https://iutbox.iut.ac.ir/index.php/s/k6CXTsHSRYM9sPj}{3(d)}
\href{https://iutbox.iut.ac.ir/index.php/s/DkbKWJsWs2raGa4}{3(e)}, and \href{https://iutbox.iut.ac.ir/index.php/s/89E8NraQ7GdfQtt}{3(f)} are available.}\label{fig4}
\end{figure*}

Based on the introduced formalism in the previous section, we generate light curves due to different DWD systems numerically. In Figure \ref{fig3}, six examples of these light curves are displayed. In each panel, three curves are plotted which show self-lensing signal (dashed green curve), eclipse by the lens disk (solid red curve), and the overall lensing/eclipsing signal (dot-dashed black curve) versus time. Also, some key parameters can be found at the top of each panel. For $\Gamma$ (linear limb-darkening coefficient of the source brightness profile), $\rho_{\star}$, $\rho_{\rm l}$, $u_{0}/\rho_{\star}$ two values are given which are due to two signals, from left to right respectively. In these panels, the signals which are specified with dark colors (dark-red and dark-green for eclipsing and self-lensing signals) happen when $M_{1}$ is the lens object and $M_{2}$ is the source star as labeled with lens$:~M_{1}$. Hence, the darker signal is scaled with $1/(1+\mathcal{F})$, and the other is scaled with $\mathcal{F}/(1+\mathcal{F})$ (see Equation \ref{magni}). We note that $\mathcal{F}$ is the ratio of the flux of the WD with the mass $M_{1}$ to the flux owing to the WD with the mass $M_{2}$ at the baseline. All light curves are simulated for the time interval $[0,~T]$ days, but the horizontal axes do not represent the full time-interval of the orbital period. Instead they show two zoomed time intervals around the lensing/eclipsing signals. For each light curve shown in one panel of Figure \ref{fig3}, we show the map of its lensing-induced images' surface brightness at a given time (which is specified with a dotted blue line on that light curve) in the corresponding panel of Figure \ref{fig4}. In each panel of this figure, three circles are shown which are the source' edge projected on lens plane (white circle), the lens's edge (black circle), and the Einstein ring (dashed red circle). The axes are normalized to the source radius projected on the lens plane ($R_{\star, \rm p}$). For each panel, an animation from similar maps versus time is available.

In Figure \ref{lca}, one of WDs has a low mass, i.e., $0.3 M_{\odot}$. When its companion WD is collinear with such a low-mass lens object as seen by the observer a complete eclipse occurs (the first signal with $\rho_{\rm l}=5$). However, since $\rho_{\star}\simeq3.8$ the lensing effect is negligible and a thick ring of images will form which is similar to the source star. We show the map of the lensing-induced images' surface brightness (by considering linear limb-darkening profile for the source star) on the lens plane at the specified time (with a blue vertical line) in Figure \ref{mca}. For this system and during the second alignment between the lens and source stars $\rho_{\star}>\rho_{\rm l}$ and the lens impact parameter is smaller than the source radius, so the eclipse happens only for some part of the images' disk. 

In the second panel, \ref{lcb}, a light curve due to an edge-on DWD is represented with one eclipse and one self-lensing. Generally, in microlensing (or self-lensing) events when $u_{0}>\rho_{\star}$ two images are formed, one inside the Einstein ring (the so-called minor image), and  the outer outside (the major image).
Here and during both alignments, $u_{0}>\rho_{\star}$ and two images form. Considering $\rho_{\rm l}$ values, minor images and some part of the major images are blocked by the lens disk. However, in the first signal, the lensing-induced shear of the major image is considerable which causes a small eclipse, in contrast to the second signal. The map of images at the given time during the second signal is shown in Figure \ref{mcb}.

As mentioned in the previous section, the light curves due to edge-on DWD systems including more massive WDs show self-lensing with either small or negligible eclipse effects. However, in these systems resulting light curves depend strongly on the lens impact parameter. In Figures \ref{lcc}, \ref{lcd}, and \ref{lce} we show the light curves due to massive DWD systems by considering three different lens impact parameters.     

In Figure \ref{lcc}, the light curve represents two self-lensing signals without considerable eclipsing effects. Because during both signals $u_{0}\gtrsim 2.5(\rho_{\star}+\rho_{\rm l})$ which means only minor images are blocked. To show this situation, we represent the images' map projected on the lens plane in Figure \ref{mcc} at the time of $44.34$ days. Accordingly the minor image is very smaller than the major image, because of the large lens impact parameter. In this light curve, the binary orbit is almost circular so that the lensing parameters for both signals are similar, although the ratio of the first signal to the second signal is $\mathcal{F}=0.13$.

In the next panel, \ref{lcd}, another light curve is plotted. In this system, $\rho_{\star}\sim 0.5$, similar to the previous light curve, but its lens impact parameters are smaller (in comparison with the previous light curve) which result larger minor images, and higher occultation effects. The occultation effects in two signals are different. In the first signal, the minor image is blocked by the lens during lensing, whereas in the second signal the minor image is not blocked by the lens disk around the time of the closest approach, because it moves out of the lens disk. We show this situation (the images' map) in Figure \ref{mcd}. For the second signal the overall flux versus time has a shorter width (although its peak does not change) because of the occultation effect.

In Figure \ref{lce}, the stellar light curve due to an edge-on DWD system including two similar WDs (with different surface temperatures) is shown. For this target, $u_{0}<\rho_{\star}$ which means the lens is crossing the source disk. Therefore, for this target whenever $u<\rho_{\star}$ the images will form a thick ring around the Einstein ring. Since $\rho_{\rm l}\simeq 0.5$, the images' ring is out of the lens disk and no occultation occurs. Hence, for transit events and while the lens is crossing the source disk, there is no occultation effect as far as $\rho_{\rm l}\lesssim \rho_{\rm in}$. We show the images' ring in Figure \ref{mce}.

In the last panel, \ref{lcf}, we display a light curve due to a common DWD system with the WDs' mass $\sim 0.5$-$0.6 M_{\odot}$. For this target $\rho_{\star}\simeq 1$ and $\rho_{\rm l}\simeq 1$. Since $u_{0}>\rho_{\star}$, images' ring does not form and instead minor and major images are formed inside and outside of the Einstein ring. Since $\rho_{\rm l}\simeq 1$, the minor image is blocked by the lens's disk always. Hence, for such events (which are common), the lensing and occultation curves are similar. The overall normalized flux versus time is a lensing-like (and symmetric) curve. The images' map due to this target at the time of the closest approach is shown in Figure \ref{mcf}.

\section{Detectability of Lensing/Eclipsing Signals from DWD Systems}\label{sec5}
In this section, we first perform Monte Carlo simulations from DWD systems, and generate their light curves as detailed in Subsection \ref{sub1}. Then, in Subsections \ref{sub3}, \ref{sub4}, and \ref{sub5} for generated light curves we simulate synthetic data points taken by the TESS, LSST, and Roman telescopes, extract the light curves with detectable lensing/eclipsing signals, and evaluate their statistics and properties.

\subsection{Monte Carlo Simulations from DWD Systems}\label{sub1}
WDs in our simulations are chosen from a big sample including $1772$ WDs within $100$ pc which were discovered from the SDSS observations \citep{2020ApJkilic}. We determine their absolute magnitudes in the TESS $T$-band, the LSST filters and the Roman W149 filter based on their relations with absolute magnitudes in the Gaia bands, i.e., $G$, $G_{\rm{RP}}$, and $G_{\rm{BP}}$ \citep{2018AJStassun,2009AJLandolt,2015ApJSAlam}. We note that the magnitude in the W149 filter is one third of the sum of absolute magnitudes in the standard filters $H$, $J$, and $K$ \citep{2017Montet,2020MNRASsajadiansalehi}. For a DWD system we need two WDs to produce a binary system. Their semi-major axis is selected from the \"{O}pik's law, i.e., a log-uniform distribution and from the range $a\in [3 R_{\star},~10^{5} R_{\star}]$. However, we exclude interacting systems which their Roche-Lobe distances \citep{1971Paczynski} are less than the source radii. The eccentricity is chosen uniformly from the range $\epsilon \in [0,~\epsilon_{\rm{max}}]$, where $\epsilon_{\rm{max}}$ is given by the known period-eccentricity correlation for binary systems \citep{2008EASMazeh}. We limit the inclination angle to $i\in [0,~5]$ degree to simulate edge-on binary systems. The projection angle $\theta$ is chosen uniformly from the range $[0,~360]$ degree. \begin{figure*}
\centering
\includegraphics[width=0.49\textwidth]{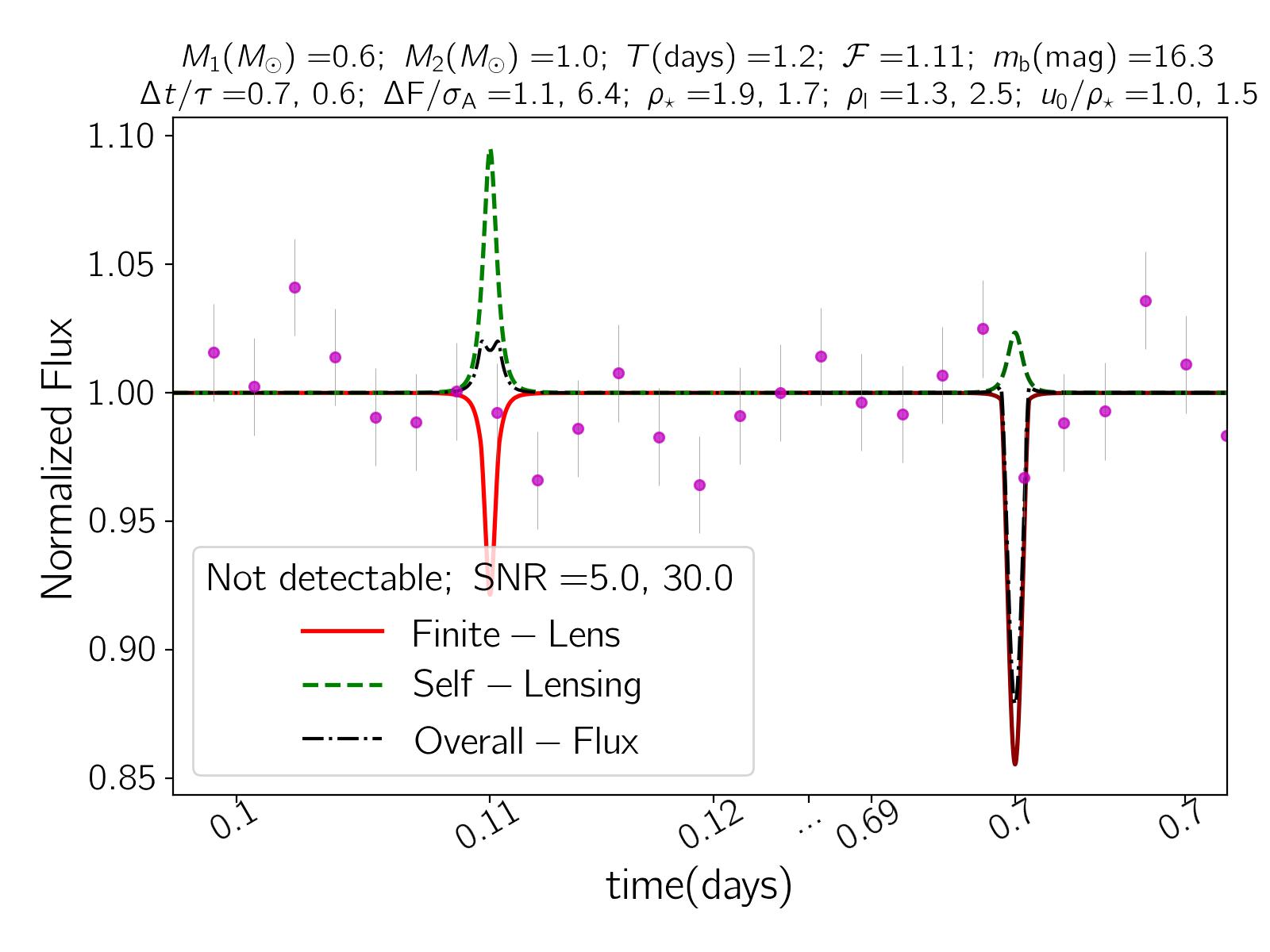}
\includegraphics[width=0.49\textwidth]{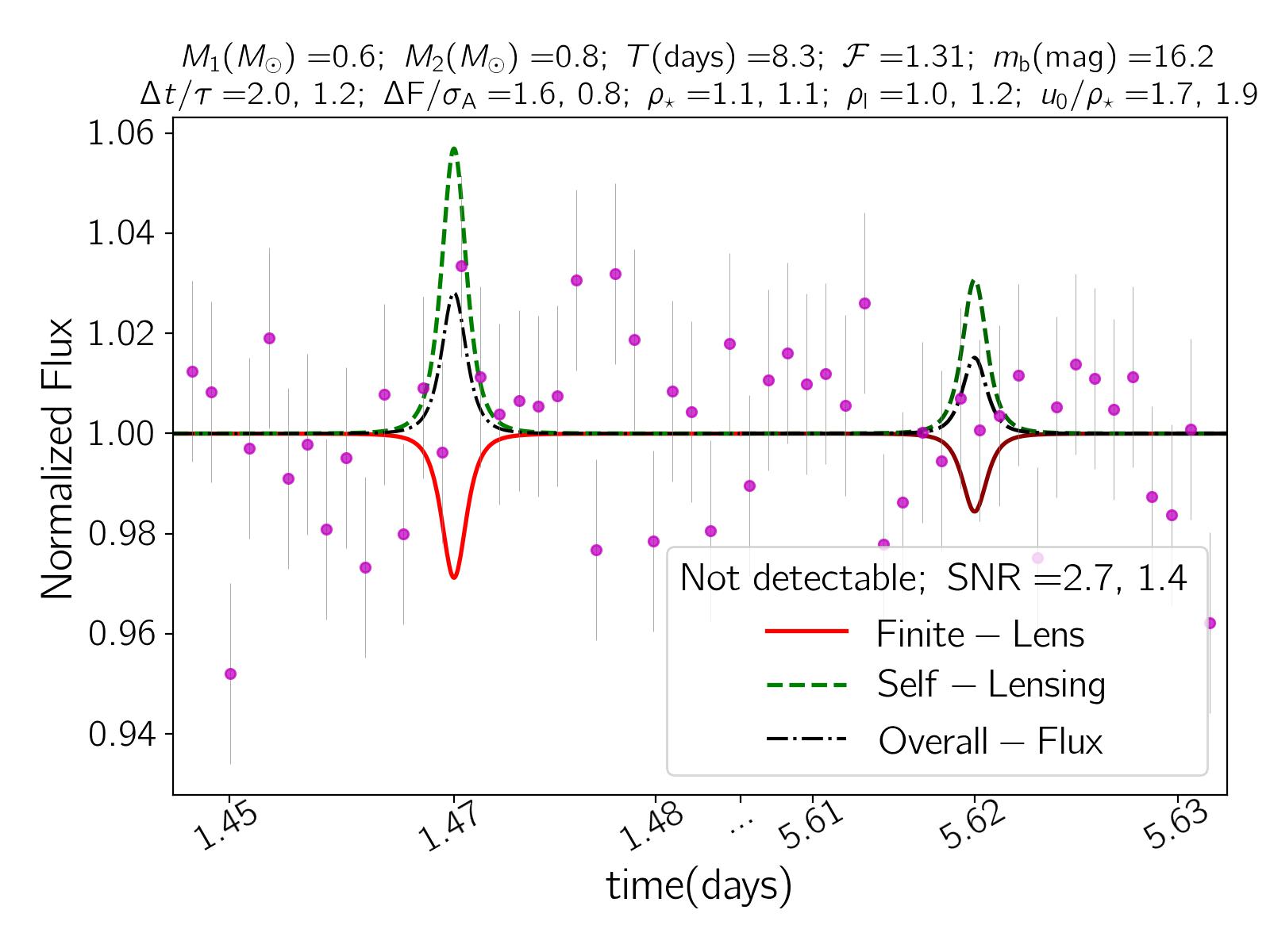}\\
\includegraphics[width=0.49\textwidth]{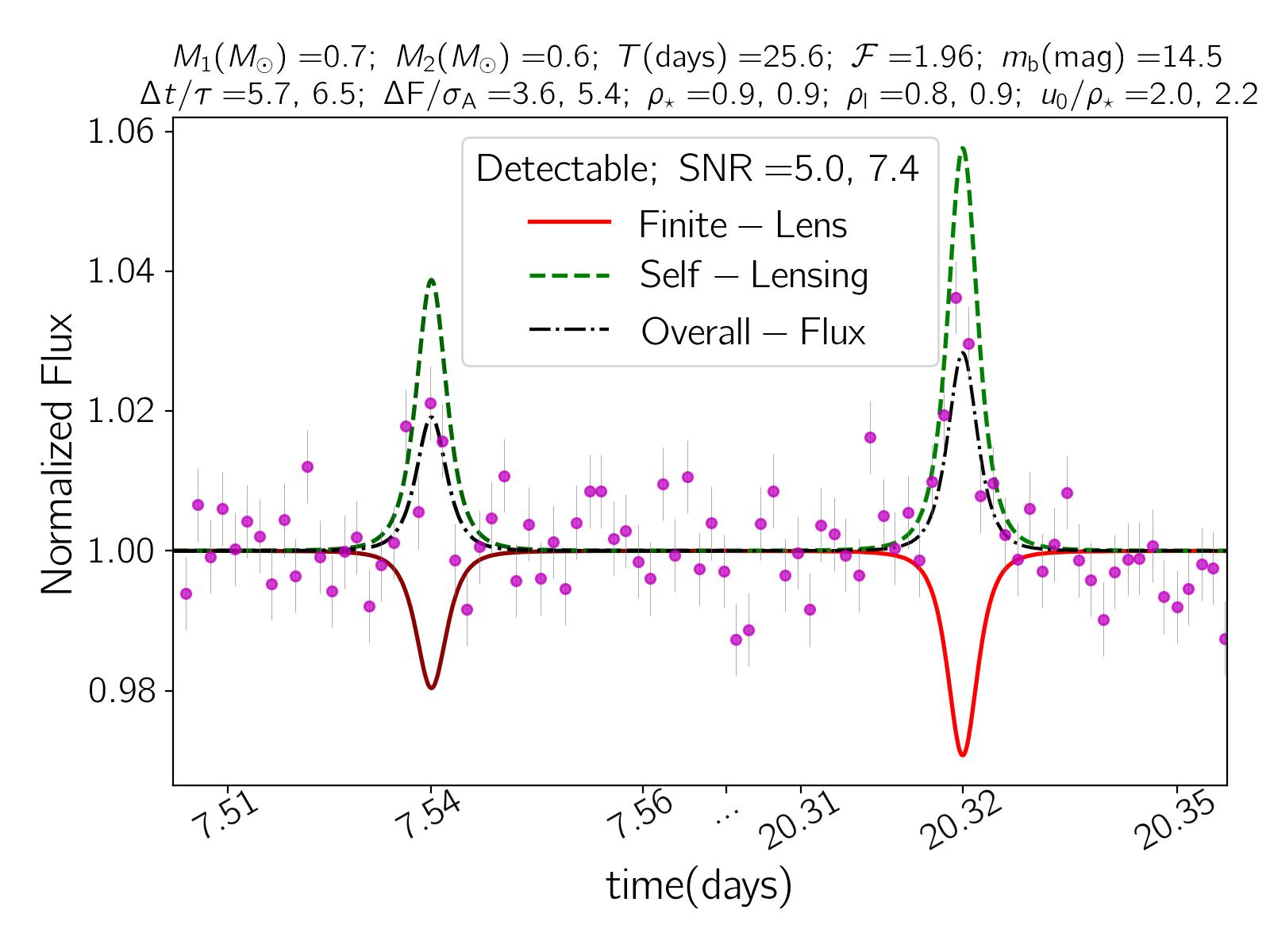}
\includegraphics[width=0.49\textwidth]{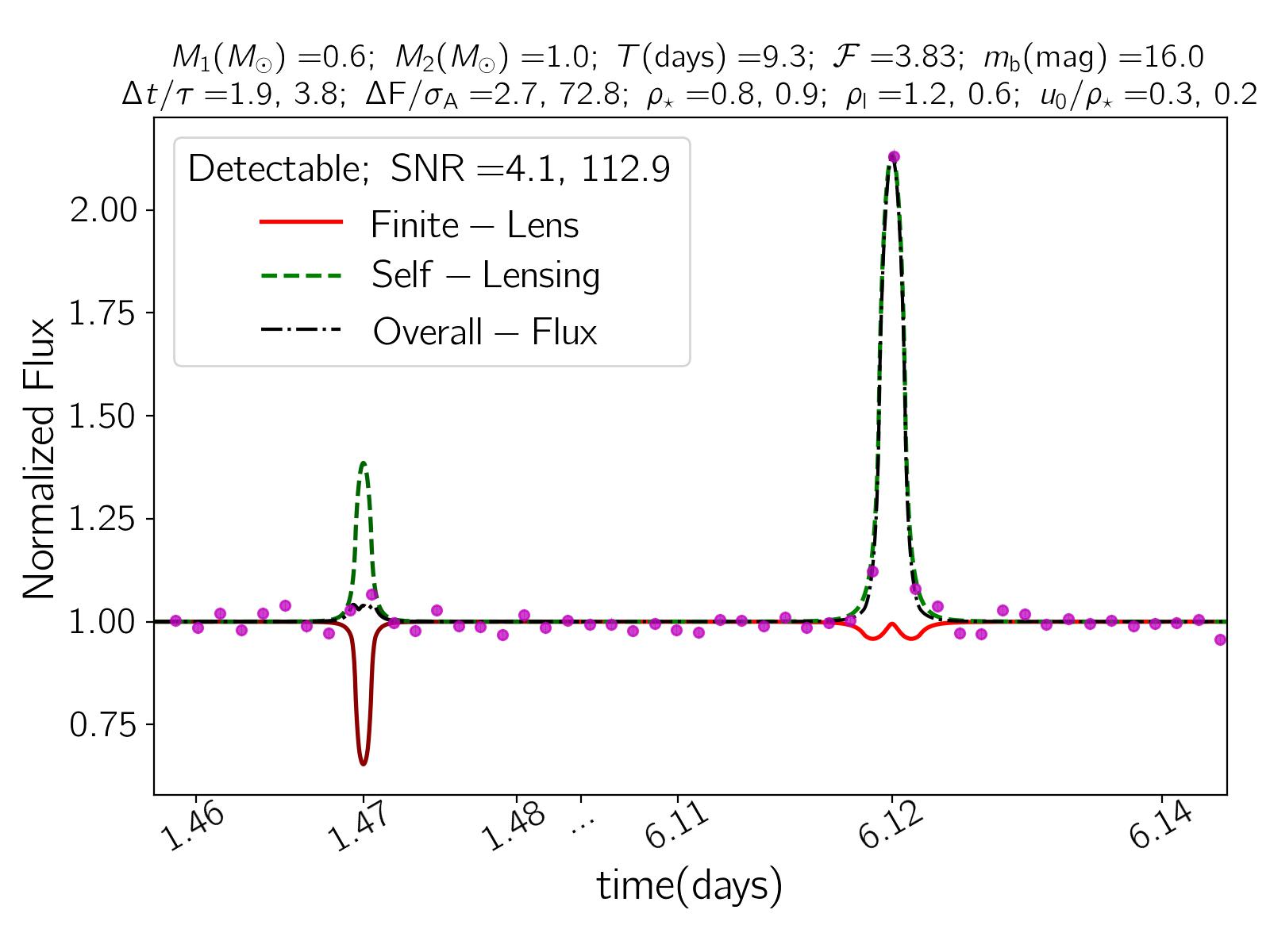}\\
\caption{Four examples of simulated light curves due to edge-on DWD systems with the synthetic data points generated based on the TESS observing strategy with a $2$-min cadence. Some physical and characteristic parameters of these light curves are mentioned at their tops. For some parameters two values are given which are related to the first and second signals, respectively.}\label{fig5}
\end{figure*}

We numerically calculate the finite-lens effect and the factor $\mathcal{O}$ using the IRS method. However, to speed up these calculations and make a large sample of all possible light curves due to different DWD systems, we improve our previous method which was explained in \citet{2024sajadianfinite}. In lensing events due to single-lens objects two generated images have the same azimuth angles with the source star, so in any step we rotate the coordinate system and set the source star on the right side of the horizontal axis ($x$-axis) at the normalized distance $u$ from the lens object, i.e., $x_{\star}=u,~y_{\star}=0$. We know that in the lensing due to a single lens object two images are formed so that the minor image is always inside the Einstein ring and the major image is close to the source position. Hence, for any given $u$ we search a rectangular part of the lens plane with the ranges $x \in \big[x_{1},~x_{2}\big]$, and $y\in \big[0,~1+2.5\rho_{\star}\big]$, where 
\begin{eqnarray}
x_{1}&=&\frac{1}{2}\big(u-\rho_{\star}-\sqrt{(u-\rho_{\star})^{2}+4}~\big)-\rho_{\star}, \nonumber\\
x_{2}&=&\frac{1}{2}\big(u+\rho_{\star}+\sqrt{(u+\rho_{\star})^{2}+4}~\big)+\rho_{\star}. 
\end{eqnarray}
However, this area covers half of the images, and the resulted $\mathcal{O}$ should be doubled. In the IRS method, we divide the Einstein radius by $200$ grids.

After simulating synthetic data points and to extract detectable events, we apply four criteria as follows. 
(i) $\rm{SNR}>3$ the signal-to-noise ratio (SNR) should be higher than $3$. Here, $\rm{SNR}=\sqrt{N_{\rm{tran}}}\Delta F/\sigma_{\rm F}$, where $\Delta F$ is the maximum lensing/eclipsing-induced depth in the normalized flux, and $N_{\rm{tran}}=T_{\rm{obs}}/T$ is the number of transits during the observing time $T_{\rm{obs}}$. $\sigma_{\rm F}\simeq \big|10^{-0.4 \sigma_{\rm m}}-1\big|$ is the photometric error in the normalized flux, and $\sigma_{\rm m}$ is the photometric error in the stellar apparent magnitude.
(ii) $N_{\rm{tran}}\ge 1$ which guarantees that at least one lensing/eclipsing signal happens during the observing time. 
(iii) $\Delta F\ge2~\sigma_{\rm F}$ which means that the depth should be deeper than 2 $\sigma_{\rm F}$.
(iv) $\Delta t\ge\tau$, where $\Delta t$ is the duration of lensing/eclipsing signals and $\tau$ is the observing cadence which is $2$ minutes for the TESS CTL observations. Two last criteria are complementary and reject some light curves with either very short or shallow signals, although they could have reasonable SNR values.

We perform three Monte Carlo simulations from these DWDs so that they are adjusted for potential observations by TESS, LSST, and Roman telescopes. Their details and results are illustrated in the following subsections, respectively.

\subsection{Detectable DWDs in the TESS Observations}\label{sub3}
In the first Monte Carlo simulation for potential observations of DWD systems by the TESS telescope we set DWD systems at the real distances (reported from the SDSS observations) of their first components. In this simulation, we simulate synthetic data points during $27.4$ days with two one-day gaps after each $12.4$-day observing time window. We note that more than $70\%$ of the TESS targets are observed during only $27.4$ days every two years (i.e., more than $70\%$ of stars inside each hemisphere are covered only by one TESS sector). We assume that these targets are in the TESS Candidate Target List (CTL, \citealp{2018AJStassun}), and are observed with a $2$-min cadence. More details of simulating the TESS observations can be found in \citet{2024sajadiankhakpash,2024sajadianAS}. The photometric error in the TESS observation is an increasing function of stellar apparent magnitude which their relation is shown in Figure (1) of \citet{2024sajadiankhakpash}. Also, CDPP (the Combined Differential Photometric Precision, \citealt{2012PASPchristiansen}) is a known metric which shows the photometric precision in normalized flux reported by the Kepler and TESS telescopes.

Four examples of simulated light curves with generated synthetic data points are represented in Figure \ref{fig5}. The structures of these plots are similar to ones shown in Figure \ref{fig3}. The magenta data points represent the synthetic data points taken by the TESS telescope. At the top of each plot, the parameters to evaluate their detectability are mentioned. For these parameters two values are given which are related to the first and second signals. For the first light curve, $\Delta t/\tau=0.7$ and no data points cover its signals, although its second signal is relatively deep with high SNR value ($30$ as mentioned in the legend part). In the second panel of Figure \ref{fig5} one light curve is shown which is not detectable as well because its signals have similar depths with the photometric errors $\Delta F/\sigma_{\rm{F}}=1.6,~0.8$. Two bottom panels display two detectable light curves.

\begin{figure}
\centering
\includegraphics[width=0.49\textwidth]{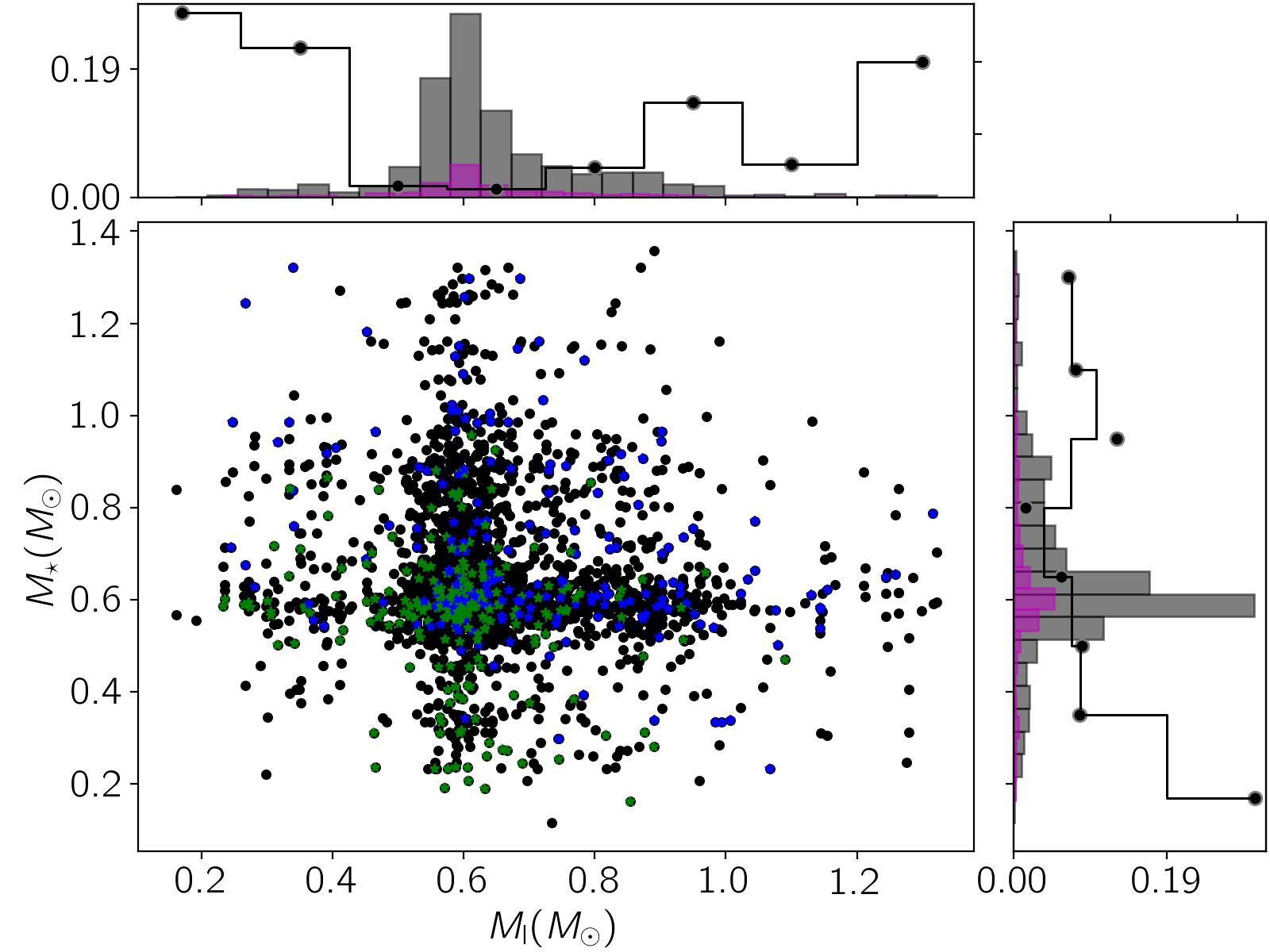}
\includegraphics[width=0.49\textwidth]{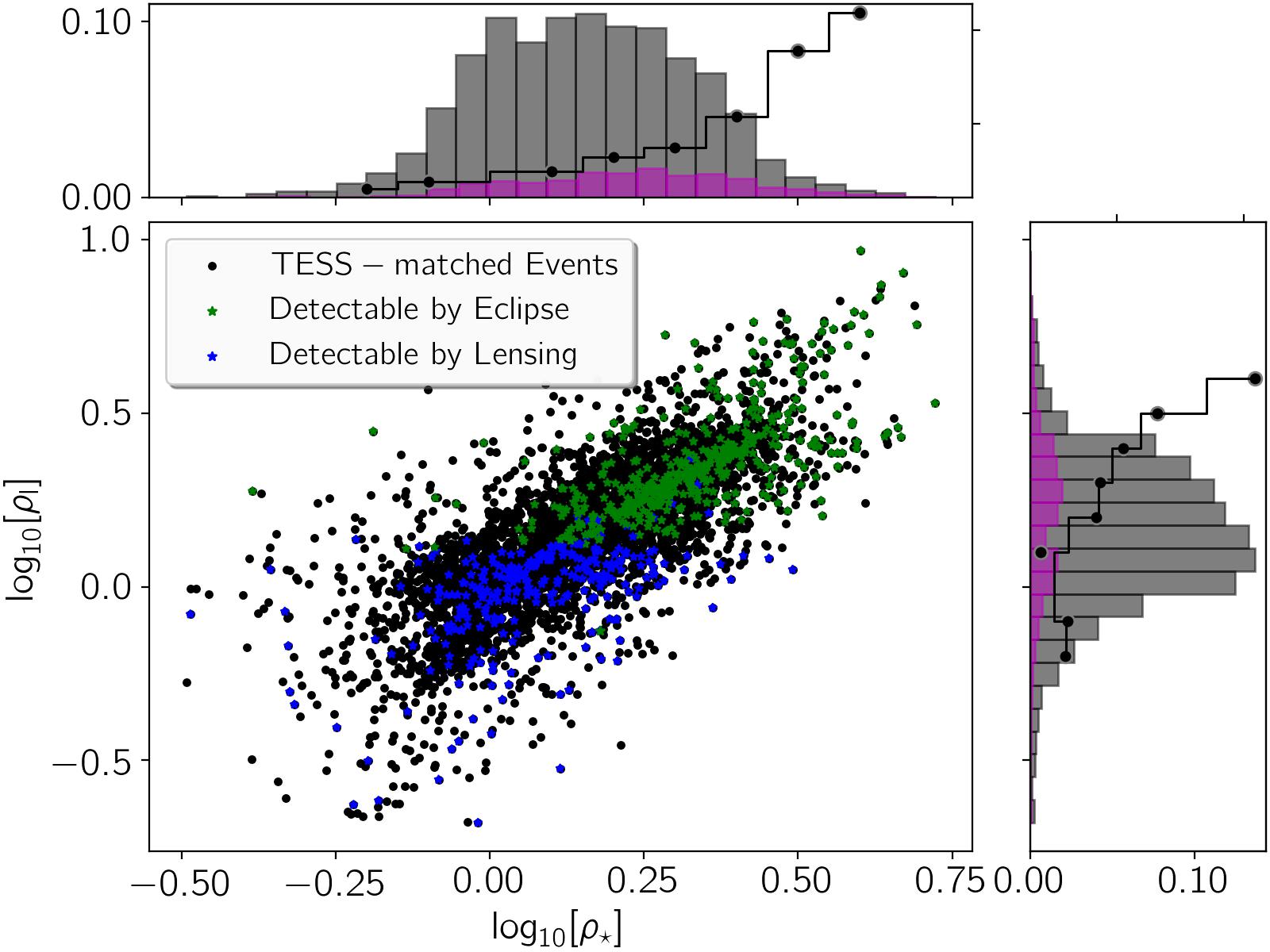}
\includegraphics[width=0.49\textwidth]{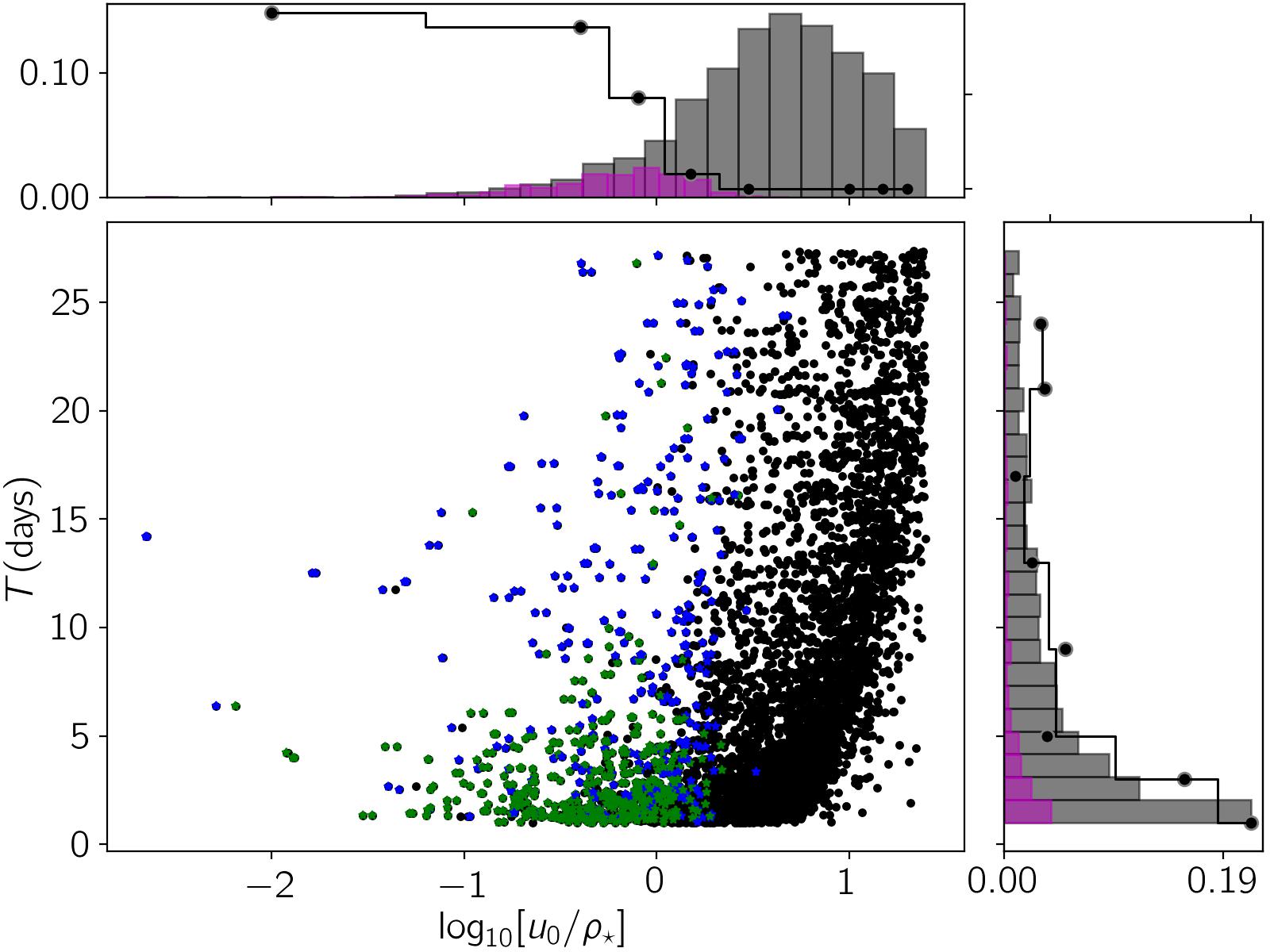}
\caption{The scatter plots of all simulated DWD systems (black circles) over three 2D spaces $M_{\rm l}(M_{\odot})-M_{\star}(M_{\odot})$ (top panel), $\log_{10}[\rho_{\star}]-\log_{10}[\rho_{\rm l}]$ (middle panel), and $\log_{10}[u_{0}/\rho_{\star}]-T(\rm{days})$ (bottom panel), with two marginal and normalized distributions. Over each plot, the targets with detectable self-lensing and eclipsing signals in the TESS observations are specified with blue and green stars, and their normalized distributions are shown with magenta color. Also, we plot the detection efficiencies over the marginal parts with black step lines. We note that in two last plots the numbers of entrances are double because each simulated light curve has two signals.}\label{fig6}
\end{figure}

In the first Monte Carlo simulation related to the TESS observations, we generate light curves and synthetic data points for events with $0.8\le T(\rm{days})\le 27.4$ and $m_{\rm{base}}\le 17.5$ mag. These events have $N_{\rm{tran}}\ge1$ and could be detectable in the TESS photometric system. Here, $m_{\rm{base}}=-2.5\log_{10}\big[10^{-0.4 m_{\rm l}}+10^{-0.4 m_{\star}}\big]+2.5\log_{10}[f_{\rm b}]$ is the baseline apparent magnitude in the TESS $T$-band due to the DWD system and blending stars. Therefore, $m_{\rm l}$, and $m_{\star}$ are the apparent magnitudes of the lens and source WDs in the TESS $T$-band. Here, for all targets, we set the blending factor $f_{\rm b}=1$ because WDs are relatively close to the observer. 

According to simulations, $\varepsilon_{1}=25.3\%$ of all simulated events have $0.8\le T(\rm{days})\le 27.4$ and $m_{\rm{base}}\le 17.5$ mag. The TESS efficiency to detect periodic lensing/eclipsing signals in the edge-on DWDs with the baseline magnitude in $T$-band less than $17.5$ mag and the orbital period $T\in [0.8,~27.4]$ days is $\varepsilon_{2}=1.53\%$. The probability that stellar orbits have the inclination angles less than $5$~degree (edge-on orbits) is $\varepsilon_{3}=5.6\%$. If we assume that the WDs are uniformly distributed in our galaxy and $14\%$ of them are in binary systems, in a spherical volume with the radius $100$ parsec around the Sun there should be $\sim 4,350$ DWD systems. And $\sim 244$ DWD systems are edge-on with $i\leq 5~\rm{deg}$ which one of them will have detectable lensing/eclipsing signals. However, the TESS telescope can detect farther and fainter WDs systems through its full frame images. Also some of the TESS CTL targets are detected longer than $27.4$ days because of overlapping some sectors. Therefore, the real number of DWD systems with detectable signals in the TESS observations should be higher.

To study the properties of detectable events, three scatter plots from simulated events are represented in Figure \ref{fig6}. In the top panel, all events with simulated light curves are shown with black circles over the 2D space $M_{\rm l}(M_{\odot})-M_{\star}(M_{\odot})$, with two marginal and normalized 1D distributions (gray ones). The DWD systems with detectable self-lensing or eclipsing signals are specified with blue and green stars, respectively. The normalized distributions due to all detectable events are represented with the magenta color. Over the marginal distributions the TESS detection efficiencies versus given parameters are offered with black step lines. The detection efficiency for a given parameter is the ratio of the number of DWD systems with that given parameter and detectable signals to total number of simulated ones with the given parameter. The first panel shows that although most of WDs have the mass $\sim 0.6 M_{\odot}$, but the detection efficiency is higher for other WD mass values. However, the detection efficiency for lower mases ($\lesssim 0.4 M_{\odot}$ for most of them complete eclipses happen) is higher than that for higher masses.  

In the middle panel of Figure \ref{fig6} we show a similar scatter plot but over the 2D space $\log_{10}[\rho_{\star}]-\log_{10}[\rho_{\rm l}]$. We note that this panel and the next one have double entrances in comparison with the top panel, because every light curve has two signals. The detection efficiency is higher for systems with large $\rho_{\star}$, and $\rho_{\rm l}$ values while most of them have eclipsing signals (specially the events with $\rho_{\rm l}\gtrsim 1$). Such systems have short orbital periods (higher $N_{\rm{tran}}$ values). The DWD systems with small $\rho_{\star}$ and $\rho_{\rm l}$ values are mostly detectable due to their self-lensing signals although their detection efficiency is small.

In the last panel we show similar scatter plot but over the 2D space $\log_{10}[u_{0}/\rho_{\star}]-T(\rm{days})$. Most of DWD systems with $u_{0}\lesssim \rho_{\star}$ have detectable signals. The systems with longer orbital periods have rather self-lensing signals than eclipsing ones. Generally, closer systems (with shorter orbital periods) have higher detection efficiencies. Hence, detection efficiency for edge-on DWD systems depends significantly on (i) orbital period, (ii) the lens impact parameters, and (iii) the total mass of the system, so that close and ELM ones with smaller inclination angles are most favorable ones with deep or even complete eclipsing signals.


\subsection{Statistics of Detectable DWDs in the LSST Observations}\label{sub4}
The observing cadence of the LSST telescope is not short enough to capture lensing/eclipsing signals in edge-on DWD systems, although its observing time window is predicted to be $10$ years. To show this point, we evaluate the duration of lensing/eclipsing signals, i.e., $\Delta t\sim 2R_{\rm{WD}}/v$, to find the longest one and study if it is realizable in the LSST data. Here, $v=2 \pi a/T$ is the relative velocity in a circular binary orbit. We consider the DWD systems with very long orbital periods $T\sim 10$ years (which is the LSST observing time span) which have the semi major axis $a \sim 10^{5}R_{\rm{WD}}$. Accordingly, the duration of their lensing/eclipsing signals would be $\sim 15~\rm{minutes}$, where we use $R_{\rm WD}\sim 0.01 R_{\sun}$. Such short signals can not be captured in the LSST observations whose cadence is three days. We note that the duration of lensing signals with small $\rho_{\star}$ values is longer and is proportional to the time of crossing the Einstein radius, the so-called Einstein crossing time $t_{\rm E}=R_{\rm E}/v\sim 1-1.5~\rm{hours}$. This time is also considerably shorter than the LSST cadence. So it seems that there is no hope that this telescope can detect lensing/eclipsing signals due to DWD systems even detached ones with long orbital periods.   


\subsection{Statistics of Detectable DWDs in the Roman Observations}\label{sub5}
We also evaluate the detectability of these signals during the Roman observations toward the Galactic bulge. This telescope will detect the Galactic bulge continuously for $62$ days with a $15$-min cadence. Its cadence and observing time span are relatively suitable to detect lensing/eclipsing signals due to DWD systems. This telescope will detect a small part of the sky toward the Galactic bulge ($\sim 2~\rm{deg}^{2}$).

\noindent Discerning isolated DWD systems in far distances is not possible (because of their faintness). For instance, a typical WD in the Galactic bulge has the average apparent magnitude $\sim 28$ mag in the Roman filter W149. Considering the stellar PSF size in the Roman observation which is $\Omega_{\rm{PSF}}=\pi\big(\rm{FWHM}/2\big)^{2}\simeq 0.09~\rm{arcs}^{2}$, the average number of blending stars whose lights enter a stellar PSF toward the Galactic bulge is $\sim 3$-$9$. Although, by adding the blending and background lights to a DWD system in the Galactic bulge it will be detectable in the Roman observation, the blending effect decreases the lensing/eclipsing signals significantly so that the resulting signals are not detectable. Hence, in the simulation we limit the distance of WDs to $D=5$ kpc from the observer and perform a similar Monte Carlo simulation. Many details for Roman-based simulations can be found in the previous papers \citep{2021MNRASsajadianhabit,2023AJsahusajadian}. 

In the simulation we make light curves for the events with $T\in [1,~62]$ days, and $m_{\rm{base}}\in [14.8,~26]$ mag in the Roman filter W149, which constitute $\varepsilon_{1}=14.84\%$ of all simulated events with the distance less than $5$ kpc. For these events, we generate the synthetic data points as taken by the Roman telescope and apply four mentioned detectability criteria to extract the detectable events. Only $\varepsilon_{2}=0.07\%$ of these events passed the criteria and had detectable lensing/eclipsing signals. To estimate the number of DWDs that this telescope can detect, we find the number of DWD systems inside a cone with the height $h= 5$ kpc and the angular area $d\Omega=2$ squared degree (its radius is $R\simeq 70$ parsec). This number can be estimated by $N_{\rm{DWD}}\simeq 0.14\times 10^{10}\times V_{\rm{R}}\big/V\simeq 26,350$ where $V_{\rm{R}}=\pi R^{2} h/3$ is the volume of the cone from our galaxy toward the Galactic bulge with the height $5 \rm{kpc}$ and the radius $R$, and $V\simeq1350 \rm{kpc}^{3}$ is the total volume of our galaxy. Considering that only $0.56\%$ of these systems are edge-on with the inclination angle less than $5$ degree, we conclude that the number of edge-on DWD systems with detectable lensing/eclipsing signals in the Roman observations is $\sim 0.1$ which is less than one.

\section{Conclusions}\label{sec6}
It was predicted that in our galaxy ten billion white dwarfs exist and $\sim9-14\%$ of them are double \citep[e.g., see][]{2017AAToonen}. Double White Dwarfs (DWDs) are interesting systems because they are progenitors of Type Ia supernovae and low-frequency gravitational waves. Detecting these systems has commonly been (and are) done through spectroscopic measurements of either their redial velocities or detecting their overlapping spectra \citep[see, e.g., ][]{2020ApJBrown,2024MNRASmunday}. If DWD systems are edge-on as seen by the observer some periodic lensing or eclipsing signals occur in their light curves and one can realize them through dense and accurate photometric observations. In this work, we studied the characterization and detection of these lensing/eclipsing signals. 

For common DWD systems three lengths of the Einstein radius, the source and lens radii have the same orders of magnitude ($\sim 0.01R_{\odot}$), which result the source and lens radii normalized to the Einstein radius ($\rho_{\star}$ and $\rho_{\rm l}$) to be mainly close to one. We considered all possible values for the mass of WDs (i.e., $[0.17,~1.4] M_{\odot}$) and a wide range for the orbital period as $T\in [1,~50]$ days, and found self-lensing signals in DWD systems enhanced with the orbital period and the masses of source and lens objects. For example, the magnification factors in edge-on DWD systems including massive WDs with $T\sim 1,~20$ days can reach $\sim6,~17$, respectively. Deep or even complete eclipsing signals will happen if the lens is an ELM WD. For instance, in most of transit events ($u<\rho_{\star}$) when $M_{\rm l} \lesssim 0.3 M_{\odot}$ complete eclipsing signals happen (see middle panels of Figure \ref{fig2}). Generally, the overall normalized flux has highest lensing-like signals for DWD systems with two massive WDs in wide orbits and deepest or complete eclipsing signals when DWD systems include one massive and one low-mass WD in close orbits so that the low-mass one is the lens object.      

For edge-on DWD systems with $\rho_{\rm l} \sim 1$, the lens object blocks the minor images versus time which results a more flattened light curve (normalized flux owing to both lensing and eclipsing effects versus time). When $\rho_{\rm l}<1$, the occultation of the minor image will not happen at the time of the closest approach and the resulting light curve will have a smaller width whereas its peak does not change because of the finite-lens effect. When $\rho_{\rm l}$ is larger than one, the eclipse effect is considerable and for $\rho_{\rm l}\gtrsim \big(u+\rho_{\star}\big)$ a complete eclipse occurs.  

We also studied detectability of lensing/eclipsing signals due to DWD systems in the TESS, LSST and Roman observations. By performing Monte Carlo simulations from these light curves and generating synthetic data points, we concluded the detection efficiency for lensing/eclipsing signals from edge-on DWD systems depends significantly on (i) orbital period, (ii) the lens impact parameters, and (iii) the mass of the lens object, so that close and ELM lenses with smaller inclination angles are most favorable ones with deep or even complete eclipsing signals. 

Inside a sphere around the Sun with the radius $100$ parsec there are $\sim 4,350$ DWD systems where $244$ of them are edge-on as seen the observer ($i\leq 5^{\circ}$). From these edge-on DWD systems, $62$ ones have the orbital periods in the range $[0.8,~27.4]~\rm{days}$ and are brighter than $17.5$ mag in the TESS $T$-band. According to our simulations, from these systems (closer than 100 parsec from us) one of them has detectable lensing/eclipsing signals in the TESS observations during a $27.4$-day observing window and with a $2$-min cadence. 

\noindent We also evaluated their detectability by the Vera C. Robin observatory and concluded that its cadence is too long to cover such short lensing/eclipsing signals even for DWD systems with the orbital periods as long as $10$ years.

\noindent Another observation in near future with a cadence relatively suitable for detecting these signals will be done with the Roman space telescope toward the Galactic bulge. This telescope surveys the Galactic bulge continuously for $62$ days with a $15$-min cadence. The Roman telescope can detect objects with the apparent magnitude $m_{\rm{W149}}\in [14.8,~26]$ mag. We tuned our simulations for observations by the Roman telescope. In this simulation we limited the observing field to seven sub-fields with the total area $2~\rm{deg}^{2}$ as will be covered by Roman \citep{2019ApJSPenny}. Considering the observing depth of this telescope we limited distances of simulated DWD systems to $5$ kpc from the observer. In this part of the sky, there should be $\sim 26,350$ DWD systems. According to our simulation, during a $62$-day observing window with a $15$-min cadence by the Roman telescope, the number of detectable lensing/eclipsing signals due to edge-on DWD systems (within an observing cone with the height $5$ kpc and the angular area $2$ squared degree) is $\sim0.1$ less than one.

\small{All simulations that have been done for this paper are available at:  \url{https://github.com/SSajadian54/Double_WhiteDwarf_LI}. Also, the codes, animations, figures and several examples of generated light curves can be found in the Zenodo repository\citep{sajadian2025Zenodo}.}\\

\bibliographystyle{aasjournal}
\bibliography{ref}{}

\begin{thebibliography}{}
\expandafter\ifx\csname natexlab\endcsname\relax\def\natexlab#1{#1}\fi
\providecommand{\url}[1]{\href{#1}{#1}}
\providecommand{\dodoi}[1]{doi:~\href{http://doi.org/#1}{\nolinkurl{#1}}}
\providecommand{\doeprint}[1]{\href{http://ascl.net/#1}{\nolinkurl{http://ascl.net/#1}}}
\providecommand{\doarXiv}[1]{\href{https://arxiv.org/abs/#1}{\nolinkurl{https://arxiv.org/abs/#1}}}

\bibitem[{{Agol}(2002)}]{2002ApJAgol}
{Agol}, E. 2002, \apj, 579, 430, \dodoi{10.1086/342880}

\bibitem[{{Agol}(2003)}]{2003ApJAgol}
---. 2003, \apj, 594, 449, \dodoi{10.1086/376833}

\bibitem[{{Alam} {et~al.}(2015){Alam}, {Albareti}, {Allende Prieto}, {Anders},
  {Anderson}, {Anderton}, {Andrews}, {Armengaud}, {Aubourg}, {Bailey}, {Basu},
  {Bautista}, {Beaton}, {Beers}, {Bender}, {Berlind}, {Beutler}, {Bhardwaj},
  {Bird}, {Bizyaev}, {Blake}, {Blanton}, {Blomqvist}, {Bochanski}, {Bolton},
  {Bovy}, {Shelden Bradley}, {Brandt}, {Brauer}, {Brinkmann}, {Brown},
  {Brownstein}, {Burden}, {Burtin}, {Busca}, {Cai}, {Capozzi}, {Carnero
  Rosell}, {Carr}, {Carrera}, {Chambers}, {Chaplin}, {Chen}, {Chiappini},
  {Chojnowski}, {Chuang}, {Clerc}, {Comparat}, {Covey}, {Croft}, {Cuesta},
  {Cunha}, {da Costa}, {Da Rio}, {Davenport}, {Dawson}, {De Lee}, {Delubac},
  {Deshpande}, {Dhital}, {Dutra-Ferreira}, {Dwelly}, {Ealet}, {Ebelke},
  {Edmondson}, {Eisenstein}, {Ellsworth}, {Elsworth}, {Epstein}, {Eracleous},
  {Escoffier}, {Esposito}, {Evans}, {Fan}, {Fern{\'a}ndez-Alvar}, {Feuillet},
  {Filiz Ak}, {Finley}, {Finoguenov}, {Flaherty}, {Fleming}, {Font-Ribera},
  {Foster}, {Frinchaboy}, {Galbraith-Frew}, {Garc{\'\i}a},
  {Garc{\'\i}a-Hern{\'a}ndez}, {Garc{\'\i}a P{\'e}rez}, {Gaulme}, {Ge},
  {G{\'e}nova-Santos}, {Georgakakis}, {Ghezzi}, {Gillespie}, {Girardi},
  {Goddard}, {Gontcho}, {Gonz{\'a}lez Hern{\'a}ndez}, {Grebel}, {Green},
  {Grieb}, {Grieves}, {Gunn}, {Guo}, {Harding}, {Hasselquist}, {Hawley},
  {Hayden}, {Hearty}, {Hekker}, {Ho}, {Hogg}, {Holley-Bockelmann}, {Holtzman},
  {Honscheid}, {Huber}, {Huehnerhoff}, {Ivans}, {Jiang}, {Johnson},
  {Kinemuchi}, {Kirkby}, {Kitaura}, {Klaene}, {Knapp}, {Kneib}, {Koenig},
  {Lam}, {Lan}, {Lang}, {Laurent}, {Le Goff}, {Leauthaud}, {Lee}, {Lee},
  {Licquia}, {Liu}, {Long}, {L{\'o}pez-Corredoira}, {Lorenzo-Oliveira},
  {Lucatello}, {Lundgren}, {Lupton}, {Mack}, {Mahadevan}, {Maia}, {Majewski},
  {Malanushenko}, {Malanushenko}, {Manchado}, {Manera}, {Mao}, {Maraston},
  {Marchwinski}, {Margala}, {Martell}, {Martig}, {Masters}, {Mathur},
  {McBride}, {McGehee}, {McGreer}, {McMahon}, {M{\'e}nard}, {Menzel},
  {Merloni}, {M{\'e}sz{\'a}ros}, {Miller}, {Miralda-Escud{\'e}}, {Miyatake},
  {Montero-Dorta}, {More}, {Morganson}, {Morice-Atkinson}, {Morrison},
  {Mosser}, {Muna}, {Myers}, {Nandra}, {Newman}, {Neyrinck}, {Nguyen},
  {Nichol}, {Nidever}, {Noterdaeme}, {Nuza}, {O'Connell}, {O'Connell},
  {O'Connell}, {Ogando}, {Olmstead}, {Oravetz}, {Oravetz}, {Osumi}, {Owen},
  {Padgett}, {Padmanabhan}, {Paegert}, {Palanque-Delabrouille}, {Pan},
  {Parejko}, {P{\^a}ris}, {Park}, {Pattarakijwanich}, {Pellejero-Ibanez},
  {Pepper}, {Percival}, {P{\'e}rez-Fournon}, {P{\'e}rez-R{\`a}fols},
  {Petitjean}, {Pieri}, {Pinsonneault}, {Porto de Mello}, {Prada}, {Prakash},
  {Price-Whelan}, {Protopapas}, {Raddick}, {Rahman}, {Reid}, {Rich}, {Rix},
  {Robin}, {Rockosi}, {Rodrigues}, {Rodr{\'\i}guez-Torres}, {Roe}, {Ross},
  {Ross}, {Rossi}, {Ruan}, {Rubi{\~n}o-Mart{\'\i}n}, {Rykoff},
  {Salazar-Albornoz}, {Salvato}, {Samushia}, {S{\'a}nchez}, {Santiago},
  {Sayres}, {Schiavon}, {Schlegel}, {Schmidt}, {Schneider}, {Schultheis},
  {Schwope}, {Sc{\'o}ccola}, {Scott}, {Sellgren}, {Seo}, {Serenelli}, {Shane},
  {Shen}, {Shetrone}, {Shu}, {Silva Aguirre}, {Sivarani}, {Skrutskie},
  {Slosar}, {Smith}, {Sobreira}, {Souto}, {Stassun}, {Steinmetz}, {Stello},
  {Strauss}, {Streblyanska}, {Suzuki}, {Swanson}, {Tan}, {Tayar}, {Terrien},
  {Thakar}, {Thomas}, {Thomas}, {Thompson}, {Tinker}, {Tojeiro}, {Troup},
  {Vargas-Maga{\~n}a}, {Vazquez}, {Verde}, {Viel}, {Vogt}, {Wake}, {Wang},
  {Weaver}, {Weinberg}, {Weiner}, {White}, {Wilson}, {Wisniewski},
  {Wood-Vasey}, {Ye`che}, {York}, {Zakamska}, {Zamora}, {Zasowski}, {Zehavi},
  {Zhao}, {Zheng}, {Zhou}, {Zhou}, {Zou}, \& {Zhu}}]{2015ApJSAlam}
{Alam}, S., {Albareti}, F.~D., {Allende Prieto}, C., {et~al.} 2015, \apjs, 219,
  12, \dodoi{10.1088/0067-0049/219/1/12}

\bibitem[{{Amaro-Seoane} {et~al.}(2017){Amaro-Seoane}, {Audley}, {Babak},
  {Baker}, {Barausse}, {Bender}, {Berti}, {Binetruy}, {Born}, {Bortoluzzi},
  {Camp}, {Caprini}, {Cardoso}, {Colpi}, {Conklin}, {Cornish}, {Cutler},
  {Danzmann}, {Dolesi}, {Ferraioli}, {Ferroni}, {Fitzsimons}, {Gair}, {Gesa
  Bote}, {Giardini}, {Gibert}, {Grimani}, {Halloin}, {Heinzel}, {Hertog},
  {Hewitson}, {Holley-Bockelmann}, {Hollington}, {Hueller}, {Inchauspe},
  {Jetzer}, {Karnesis}, {Killow}, {Klein}, {Klipstein}, {Korsakova}, {Larson},
  {Livas}, {Lloro}, {Man}, {Mance}, {Martino}, {Mateos}, {McKenzie},
  {McWilliams}, {Miller}, {Mueller}, {Nardini}, {Nelemans}, {Nofrarias},
  {Petiteau}, {Pivato}, {Plagnol}, {Porter}, {Reiche}, {Robertson},
  {Robertson}, {Rossi}, {Russano}, {Schutz}, {Sesana}, {Shoemaker}, {Slutsky},
  {Sopuerta}, {Sumner}, {Tamanini}, {Thorpe}, {Troebs}, {Vallisneri},
  {Vecchio}, {Vetrugno}, {Vitale}, {Volonteri}, {Wanner}, {Ward}, {Wass},
  {Weber}, {Ziemer}, \& {Zweifel}}]{2017AmaroSeoane}
{Amaro-Seoane}, P., {Audley}, H., {Babak}, S., {et~al.} 2017, arXiv e-prints,
  arXiv:1702.00786, \dodoi{10.48550/arXiv.1702.00786}

\bibitem[{{Ambrosino}(2020)}]{2020Federico}
{Ambrosino}, F. 2020, arXiv e-prints, arXiv:2012.01242,
  \dodoi{10.48550/arXiv.2012.01242}

\bibitem[{{Bellm} {et~al.}(2019){Bellm}, {Kulkarni}, {Graham}, {Dekany},
  {Smith}, {Riddle}, {Masci}, {Helou}, {Prince}, {Adams}, {Barbarino},
  {Barlow}, {Bauer}, {Beck}, {Belicki}, {Biswas}, {Blagorodnova}, {Bodewits},
  {Bolin}, {Brinnel}, {Brooke}, {Bue}, {Bulla}, {Burruss}, {Cenko}, {Chang},
  {Connolly}, {Coughlin}, {Cromer}, {Cunningham}, {De}, {Delacroix}, {Desai},
  {Duev}, {Eadie}, {Farnham}, {Feeney}, {Feindt}, {Flynn}, {Franckowiak},
  {Frederick}, {Fremling}, {Gal-Yam}, {Gezari}, {Giomi}, {Goldstein},
  {Golkhou}, {Goobar}, {Groom}, {Hacopians}, {Hale}, {Henning}, {Ho}, {Hover},
  {Howell}, {Hung}, {Huppenkothen}, {Imel}, {Ip}, {Ivezi{\'c}}, {Jackson},
  {Jones}, {Juric}, {Kasliwal}, {Kaspi}, {Kaye}, {Kelley}, {Kowalski},
  {Kramer}, {Kupfer}, {Landry}, {Laher}, {Lee}, {Lin}, {Lin}, {Lunnan},
  {Giomi}, {Mahabal}, {Mao}, {Miller}, {Monkewitz}, {Murphy}, {Ngeow},
  {Nordin}, {Nugent}, {Ofek}, {Patterson}, {Penprase}, {Porter}, {Rauch},
  {Rebbapragada}, {Reiley}, {Rigault}, {Rodriguez}, {van Roestel}, {Rusholme},
  {van Santen}, {Schulze}, {Shupe}, {Singer}, {Soumagnac}, {Stein}, {Surace},
  {Sollerman}, {Szkody}, {Taddia}, {Terek}, {Van Sistine}, {van Velzen},
  {Vestrand}, {Walters}, {Ward}, {Ye}, {Yu}, {Yan}, \&
  {Zolkower}}]{2019PASPbellem}
{Bellm}, E.~C., {Kulkarni}, S.~R., {Graham}, M.~J., {et~al.} 2019, \pasp, 131,
  018002, \dodoi{10.1088/1538-3873/aaecbe}

\bibitem[{{Beskin} \& {Tuntsov}(2002)}]{2002AABeskin}
{Beskin}, G.~M., \& {Tuntsov}, A.~V. 2002, \aap, 394, 489,
  \dodoi{10.1051/0004-6361:20021150}

\bibitem[{{Bozza}(2010)}]{2010MNRASBozza}
{Bozza}, V. 2010, \mnras, 408, 2188, \dodoi{10.1111/j.1365-2966.2010.17265.x}

\bibitem[{{Bozza} {et~al.}(2018){Bozza}, {Bachelet}, {Bartoli{\'c}}, {Heintz},
  {Hoag}, \& {Hundertmark}}]{2018MNRABozza}
{Bozza}, V., {Bachelet}, E., {Bartoli{\'c}}, F., {et~al.} 2018, \mnras, 479,
  5157, \dodoi{10.1093/mnras/sty1791}

\bibitem[{{Brown} {et~al.}(2010){Brown}, {Kilic}, {Allende Prieto}, \&
  {Kenyon}}]{2010ApJELTsurvey}
{Brown}, W.~R., {Kilic}, M., {Allende Prieto}, C., \& {Kenyon}, S.~J. 2010,
  \apj, 723, 1072, \dodoi{10.1088/0004-637X/723/2/1072}

\bibitem[{{Brown} {et~al.}(2020){Brown}, {Kilic}, {Kosakowski}, {Andrews},
  {Heinke}, {Ag{\"u}eros}, {Camilo}, {Gianninas}, {Hermes}, \&
  {Kenyon}}]{2020ApJBrown}
{Brown}, W.~R., {Kilic}, M., {Kosakowski}, A., {et~al.} 2020, \apj, 889, 49,
  \dodoi{10.3847/1538-4357/ab63cd}

\bibitem[{{Burdge} {et~al.}(2019){Burdge}, {Fuller}, {Phinney}, {van Roestel},
  {Claret}, {Cukanovaite}, {Gentile Fusillo}, {Coughlin}, {Kaplan}, {Kupfer},
  {Tremblay}, {Dekany}, {Duev}, {Feeney}, {Riddle}, {Kulkarni}, \&
  {Prince}}]{2019ApJBurdge}
{Burdge}, K.~B., {Fuller}, J., {Phinney}, E.~S., {et~al.} 2019, \apjl, 886,
  L12, \dodoi{10.3847/2041-8213/ab53e5}

\bibitem[{{Chen} {et~al.}(2022){Chen}, {Tauris}, {Chen}, \&
  {Han}}]{2022ApJChen}
{Chen}, H.-L., {Tauris}, T.~M., {Chen}, X., \& {Han}, Z. 2022, \apj, 925, 89,
  \dodoi{10.3847/1538-4357/ac3bb6}

\bibitem[{{Christiansen} {et~al.}(2012){Christiansen}, {Jenkins}, {Caldwell},
  {Burke}, {Tenenbaum}, {Seader}, {Thompson}, {Barclay}, {Clarke}, {Li},
  {Smith}, {Stumpe}, {Twicken}, \& {Van Cleve}}]{2012PASPchristiansen}
{Christiansen}, J.~L., {Jenkins}, J.~M., {Caldwell}, D.~A., {et~al.} 2012,
  \pasp, 124, 1279, \dodoi{10.1086/668847}

\bibitem[{{Garza Navarro} \& {Wilson}(2021)}]{2021GrazaWilson}
{Garza Navarro}, N., \& {Wilson}, D.~J. 2021, Research Notes of the American
  Astronomical Society, 5, 269, \dodoi{10.3847/2515-5172/ac3af7}

\bibitem[{{Gould}(1995)}]{1995ApJGould}
{Gould}, A. 1995, \apj, 446, 541, \dodoi{10.1086/175812}

\bibitem[{{Gould} \& {Gaucherel}(1996)}]{1996Gould}
{Gould}, A., \& {Gaucherel}, C. 1996, arXiv e-prints, astro,
  \dodoi{10.48550/arXiv.astro-ph/9606105}

\bibitem[{{Han}(2016)}]{2016ApJHan}
{Han}, C. 2016, \apj, 820, 53, \dodoi{10.3847/0004-637X/820/1/53}

\bibitem[{{Jin} {et~al.}(2024){Jin}, {Ma}, {Shao}, \& {Wang}}]{2024arXivJin}
{Jin}, H.-M., {Ma}, B., {Shao}, Y., \& {Wang}, Y. 2024, arXiv e-prints,
  arXiv:2406.16474, \dodoi{10.48550/arXiv.2406.16474}

\bibitem[{{Kawahara} {et~al.}(2018){Kawahara}, {Masuda}, {MacLeod}, {Latham},
  {Bieryla}, \& {Benomar}}]{2018AJKawahara}
{Kawahara}, H., {Masuda}, K., {MacLeod}, M., {et~al.} 2018, \aj, 155, 144,
  \dodoi{10.3847/1538-3881/aaaaaf}

\bibitem[{{Kilic} {et~al.}(2020{\natexlab{a}}){Kilic}, {B{\'e}dard},
  {Bergeron}, \& {Kosakowski}}]{2020kilicmnras}
{Kilic}, M., {B{\'e}dard}, A., {Bergeron}, P., \& {Kosakowski}, A.
  2020{\natexlab{a}}, \mnras, 493, 2805, \dodoi{10.1093/mnras/staa466}

\bibitem[{{Kilic} {et~al.}(2020{\natexlab{b}}){Kilic}, {Bergeron},
  {Kosakowski}, {Brown}, {Ag{\"u}eros}, \& {Blouin}}]{2020ApJkilic}
{Kilic}, M., {Bergeron}, P., {Kosakowski}, A., {et~al.} 2020{\natexlab{b}},
  \apj, 898, 84, \dodoi{10.3847/1538-4357/ab9b8d}

\bibitem[{{Kruse} \& {Agol}(2014)}]{KruseAgol2014}
{Kruse}, E., \& {Agol}, E. 2014, Science, 344, 275,
  \dodoi{10.1126/science.1251999}

\bibitem[{{Landolt}(2009)}]{2009AJLandolt}
{Landolt}, A.~U. 2009, \aj, 137, 4186, \dodoi{10.1088/0004-6256/137/5/4186}

\bibitem[{{Li} {et~al.}(2019){Li}, {Chen}, {Chen}, \& {Han}}]{2019ApJLi}
{Li}, Z., {Chen}, X., {Chen}, H.-L., \& {Han}, Z. 2019, \apj, 871, 148,
  \dodoi{10.3847/1538-4357/aaf9a1}

\bibitem[{{LSST Science Collaboration} {et~al.}(2009){LSST Science
  Collaboration}, {Abell}, {Allison}, {Anderson}, {Andrew}, {Angel}, {Armus},
  {Arnett}, {Asztalos}, {Axelrod}, {Bailey}, {Ballantyne}, {Bankert},
  {Barkhouse}, {Barr}, {Barrientos}, {Barth}, {Bartlett}, {Becker}, {Becla},
  {Beers}, {Bernstein}, {Biswas}, {Blanton}, {Bloom}, {Bochanski}, {Boeshaar},
  {Borne}, {Bradac}, {Brandt}, {Bridge}, {Brown}, {Brunner}, {Bullock},
  {Burgasser}, {Burge}, {Burke}, {Cargile}, {Chandrasekharan}, {Chartas},
  {Chesley}, {Chu}, {Cinabro}, {Claire}, {Claver}, {Clowe}, {Connolly}, {Cook},
  {Cooke}, {Cooray}, {Covey}, {Culliton}, {de Jong}, {de Vries}, {Debattista},
  {Delgado}, {Dell'Antonio}, {Dhital}, {Di Stefano}, {Dickinson}, {Dilday},
  {Djorgovski}, {Dobler}, {Donalek}, {Dubois-Felsmann}, {Durech},
  {Eliasdottir}, {Eracleous}, {Eyer}, {Falco}, {Fan}, {Fassnacht}, {Ferguson},
  {Fernandez}, {Fields}, {Finkbeiner}, {Figueroa}, {Fox}, {Francke}, {Frank},
  {Frieman}, {Fromenteau}, {Furqan}, {Galaz}, {Gal-Yam}, {Garnavich},
  {Gawiser}, {Geary}, {Gee}, {Gibson}, {Gilmore}, {Grace}, {Green}, {Gressler},
  {Grillmair}, {Habib}, {Haggerty}, {Hamuy}, {Harris}, {Hawley}, {Heavens},
  {Hebb}, {Henry}, {Hileman}, {Hilton}, {Hoadley}, {Holberg}, {Holman},
  {Howell}, {Infante}, {Ivezic}, {Jacoby}, {Jain}, {R}, {Jedicke}, {Jee},
  {Garrett Jernigan}, {Jha}, {Johnston}, {Jones}, {Juric}, {Kaasalainen},
  {Styliani}, {Kafka}, {Kahn}, {Kaib}, {Kalirai}, {Kantor}, {Kasliwal},
  {Keeton}, {Kessler}, {Knezevic}, {Kowalski}, {Krabbendam}, {Krughoff},
  {Kulkarni}, {Kuhlman}, {Lacy}, {Lepine}, {Liang}, {Lien}, {Lira}, {Long},
  {Lorenz}, {Lotz}, {Lupton}, {Lutz}, {Macri}, {Mahabal}, {Mandelbaum},
  {Marshall}, {May}, {McGehee}, {Meadows}, {Meert}, {Milani}, {Miller},
  {Miller}, {Mills}, {Minniti}, {Monet}, {Mukadam}, {Nakar}, {Neill}, {Newman},
  {Nikolaev}, {Nordby}, {O'Connor}, {Oguri}, {Oliver}, {Olivier}, {Olsen},
  {Olsen}, {Olszewski}, {Oluseyi}, {Padilla}, {Parker}, {Pepper}, {Peterson},
  {Petry}, {Pinto}, {Pizagno}, {Popescu}, {Prsa}, {Radcka}, {Raddick},
  {Rasmussen}, {Rau}, {Rho}, {Rhoads}, {Richards}, {Ridgway}, {Robertson},
  {Roskar}, {Saha}, {Sarajedini}, {Scannapieco}, {Schalk}, {Schindler},
  {Schmidt}, {Schmidt}, {Schneider}, {Schumacher}, {Scranton}, {Sebag},
  {Seppala}, {Shemmer}, {Simon}, {Sivertz}, {Smith}, {Allyn Smith}, {Smith},
  {Spitz}, {Stanford}, {Stassun}, {Strader}, {Strauss}, {Stubbs}, {Sweeney},
  {Szalay}, {Szkody}, {Takada}, {Thorman}, {Trilling}, {Trimble}, {Tyson}, {Van
  Berg}, {Vanden Berk}, {VanderPlas}, {Verde}, {Vrsnak}, {Walkowicz},
  {Wandelt}, {Wang}, {Wang}, {Warner}, {Wechsler}, {West}, {Wiecha},
  {Williams}, {Willman}, {Wittman}, {Wolff}, {Wood-Vasey}, {Wozniak}, {Young},
  {Zentner}, \& {Zhan}}]{2009LSSTbook}
{LSST Science Collaboration}, {Abell}, P.~A., {Allison}, J., {et~al.} 2009,
  arXiv e-prints, arXiv:0912.0201, \dodoi{10.48550/arXiv.0912.0201}

\bibitem[{{Luo} {et~al.}(2016){Luo}, {Chen}, {Duan}, {Gong}, {Hu}, {Ji}, {Liu},
  {Mei}, {Milyukov}, {Sazhin}, {Shao}, {Toth}, {Tu}, {Wang}, {Wang}, {Yeh},
  {Zhan}, {Zhang}, {Zharov}, \& {Zhou}}]{2016Luo}
{Luo}, J., {Chen}, L.-S., {Duan}, H.-Z., {et~al.} 2016, Classical and Quantum
  Gravity, 33, 035010, \dodoi{10.1088/0264-9381/33/3/035010}

\bibitem[{{Maeder}(1973{\natexlab{a}})}]{1973AAMaeder}
{Maeder}, A. 1973{\natexlab{a}}, \aap, 26, 215

\bibitem[{{Maeder}(1973{\natexlab{b}})}]{1973Maeder}
---. 1973{\natexlab{b}}, \aap, 26, 215

\bibitem[{{Masuda} {et~al.}(2019){Masuda}, {Kawahara}, {Latham}, {Bieryla},
  {Kunitomo}, {MacLeod}, \& {Aoki}}]{2019ApJLMasuda}
{Masuda}, K., {Kawahara}, H., {Latham}, D.~W., {et~al.} 2019, \apjl, 881, L3,
  \dodoi{10.3847/2041-8213/ab321b}

\bibitem[{{Mazeh}(2008)}]{2008EASMazeh}
{Mazeh}, T. 2008, in EAS Publications Series, Vol.~29, EAS Publications Series,
  ed. M.~J. {Goupil} \& J.~P. {Zahn}, 1--65, \dodoi{10.1051/eas:0829001}

\bibitem[{{Montet} {et~al.}(2017){Montet}, {Yee}, \& {Penny}}]{2017Montet}
{Montet}, B.~T., {Yee}, J.~C., \& {Penny}, M.~T. 2017, \pasp, 129, 044401,
  \dodoi{10.1088/1538-3873/aa57fb}

\bibitem[{{Munday} {et~al.}(2023){Munday}, {Tremblay}, {Hermes}, {Barlow},
  {Pelisoli}, {Marsh}, {Parsons}, {Jones}, {Kepler}, {Brown}, {Littlefair},
  {Hegedus}, {Baran}, {Breedt}, {Dhillon}, {Dyer}, {Green}, {Kennedy}, {Kerry},
  {Lopez}, {Romero}, {Sahman}, \& {Worters}}]{2023MNRASMunday}
{Munday}, J., {Tremblay}, P.~E., {Hermes}, J.~J., {et~al.} 2023, \mnras, 525,
  1814, \dodoi{10.1093/mnras/stad2347}

\bibitem[{{Munday} {et~al.}(2024){Munday}, {Pelisoli}, {Tremblay}, {Marsh},
  {Nelemans}, {B{\'e}dard}, {Toonen}, {Breedt}, {Cunningham}, {O'Brien}, \&
  {Dawson}}]{2024MNRASmunday}
{Munday}, J., {Pelisoli}, I., {Tremblay}, P.~E., {et~al.} 2024, \mnras, 532,
  2534, \dodoi{10.1093/mnras/stae1645}

\bibitem[{{Nauenberg}(1972)}]{1972ApJWDMR}
{Nauenberg}, M. 1972, \apj, 175, 417, \dodoi{10.1086/151568}

\bibitem[{{Nelemans} {et~al.}(2001){Nelemans}, {Portegies Zwart}, {Verbunt}, \&
  {Yungelson}}]{2001Nelemans}
{Nelemans}, G., {Portegies Zwart}, S.~F., {Verbunt}, F., \& {Yungelson}, L.~R.
  2001, \aap, 368, 939, \dodoi{10.1051/0004-6361:20010049}

\bibitem[{{Nir} \& {Bloom}(2023)}]{2023arXivNir}
{Nir}, G., \& {Bloom}, J.~S. 2023, arXiv e-prints, arXiv:2311.14392,
  \dodoi{10.48550/arXiv.2311.14392}

\bibitem[{{Paczy{\'n}ski}(1971)}]{1971Paczynski}
{Paczy{\'n}ski}, B. 1971, \araa, 9, 183,
  \dodoi{10.1146/annurev.aa.09.090171.001151}

\bibitem[{{Penny} {et~al.}(2019){Penny}, {Gaudi}, {Kerins}, {Rattenbury},
  {Mao}, {Robin}, \& {Calchi Novati}}]{2019ApJSPenny}
{Penny}, M.~T., {Gaudi}, B.~S., {Kerins}, E., {et~al.} 2019, \apjs, 241, 3,
  \dodoi{10.3847/1538-4365/aafb69}

\bibitem[{{Qin} {et~al.}(1997){Qin}, {Wu}, \& {Zou}}]{1997ChPhLQin}
{Qin}, B., {Wu}, X.-p., \& {Zou}, Z.-l. 1997, Chinese Physics Letters, 14, 155,
  \dodoi{10.1088/0256-307X/14/2/022}

\bibitem[{{Ricker} {et~al.}(2014){Ricker}, {Winn}, {Vanderspek}, {Latham},
  {Bakos}, {Bean}, {Berta-Thompson}, {Brown}, {Buchhave}, {Butler}, {Butler},
  {Chaplin}, {Charbonneau}, {Christensen-Dalsgaard}, {Clampin}, {Deming},
  {Doty}, {De Lee}, {Dressing}, {Dunham}, {Endl}, {Fressin}, {Ge}, {Henning},
  {Holman}, {Howard}, {Ida}, {Jenkins}, {Jernigan}, {Johnson}, {Kaltenegger},
  {Kawai}, {Kjeldsen}, {Laughlin}, {Levine}, {Lin}, {Lissauer}, {MacQueen},
  {Marcy}, {McCullough}, {Morton}, {Narita}, {Paegert}, {Palle}, {Pepe},
  {Pepper}, {Quirrenbach}, {Rinehart}, {Sasselov}, {Sato}, {Seager},
  {Sozzetti}, {Stassun}, {Sullivan}, {Szentgyorgyi}, {Torres}, {Udry}, \&
  {Villasenor}}]{Ricker2024}
{Ricker}, G.~R., {Winn}, J.~N., {Vanderspek}, R., {et~al.} 2014, in Society of
  Photo-Optical Instrumentation Engineers (SPIE) Conference Series, Vol. 9143,
  Space Telescopes and Instrumentation 2014: Optical, Infrared, and Millimeter
  Wave, ed. J.~{Oschmann}, Jacobus~M., M.~{Clampin}, G.~G. {Fazio}, \& H.~A.
  {MacEwen}, 914320, \dodoi{10.1117/12.2063489}

\bibitem[{Ruan {et~al.}(2020)Ruan, Guo, Cai, \& Zhang}]{Ruan2020}
Ruan, W.-H., Guo, Z.-K., Cai, R.-G., \& Zhang, Y.-Z. 2020, International
  Journal of Modern Physics A, 35, 2050075, \dodoi{10.1142/S0217751X2050075X}

\bibitem[{{Sajadian}(2014)}]{2014MNRASorbit}
{Sajadian}, S. 2014, \mnras, 439, 3007, \dodoi{10.1093/mnras/stu158}

\bibitem[{{Sajadian}(2021)}]{2021MNRASsajadianhabit}
---. 2021, \mnras, 508, 5991, \dodoi{10.1093/mnras/stab2942}

\bibitem[{{Sajadian}(2023)}]{2023sajadiandegeneracy}
---. 2023, \mnras, 521, 6383, \dodoi{10.1093/mnras/stad945}

\bibitem[{sajadian(2025)}]{sajadian2025Zenodo}
sajadian, s. 2025, Software, figures and animations for simulating
  lensing/eclipsing in double white dwarf systems,  Zenodo,
  \dodoi{10.5281/zenodo.14631446}

\bibitem[{{Sajadian} \& {Afshordi}(2024)}]{2024sajadianAS}
{Sajadian}, S., \& {Afshordi}, N. 2024, arXiv e-prints, arXiv:2409.12441,
  \dodoi{10.48550/arXiv.2409.12441}

\bibitem[{{Sajadian} \& {Fatheddin}(2024)}]{2024sajadianfinite}
{Sajadian}, S., \& {Fatheddin}, H. 2024, arXiv e-prints, arXiv:2410.04550,
  \dodoi{10.48550/arXiv.2410.04550}

\bibitem[{{Sajadian} {et~al.}(2024){Sajadian}, {Kalantari}, {Fatheddin}, \&
  {Khakpash}}]{2024sajadiankhakpash}
{Sajadian}, S., {Kalantari}, A., {Fatheddin}, H., \& {Khakpash}, S. 2024, arXiv
  e-prints, arXiv:2408.14231, \dodoi{10.48550/arXiv.2408.14231}

\bibitem[{{Sajadian} \& {Rahvar}(2010)}]{2010MNRAsajadian}
{Sajadian}, S., \& {Rahvar}, S. 2010, \mnras, 407, 373,
  \dodoi{10.1111/j.1365-2966.2010.16901.x}

\bibitem[{{Sajadian} \& {Sahu}(2023)}]{2023AJsahusajadian}
{Sajadian}, S., \& {Sahu}, K.~C. 2023, \aj, 165, 96,
  \dodoi{10.3847/1538-3881/acb20f}

\bibitem[{{Sajadian} \& {Salehi}(2020)}]{2020MNRASsajadiansalehi}
{Sajadian}, S., \& {Salehi}, A. 2020, \mnras, 498, 1298,
  \dodoi{10.1093/mnras/staa2377}

\bibitem[{{Shporer} {et~al.}(2010){Shporer}, {Kaplan}, {Steinfadt}, {Bildsten},
  {Howell}, \& {Mazeh}}]{2010ApJShoperer}
{Shporer}, A., {Kaplan}, D.~L., {Steinfadt}, J. D.~R., {et~al.} 2010, \apjl,
  725, L200, \dodoi{10.1088/2041-8205/725/2/L200}

\bibitem[{{Soethe} \& {Kepler}(2021)}]{2021Soethe}
{Soethe}, L.~T.~T., \& {Kepler}, S.~O. 2021, \mnras, 506, 3266,
  \dodoi{10.1093/mnras/stab1916}

\bibitem[{{Sorabella} {et~al.}(2024){Sorabella}, {Laycock}, {Christodoulou}, \&
  {Bhattacharya}}]{2024ApJSorebella}
{Sorabella}, N.~M., {Laycock}, S. G.~T., {Christodoulou}, D.~M., \&
  {Bhattacharya}, S. 2024, \apjl, 961, L45, \dodoi{10.3847/2041-8213/ad19dc}

\bibitem[{{Stassun} {et~al.}(2018){Stassun}, {Oelkers}, {Pepper}, {Paegert},
  {De Lee}, {Torres}, {Latham}, {Charpinet}, {Dressing}, {Huber}, {Kane},
  {L{\'e}pine}, {Mann}, {Muirhead}, {Rojas-Ayala}, {Silvotti}, {Fleming},
  {Levine}, \& {Plavchan}}]{2018AJStassun}
{Stassun}, K.~G., {Oelkers}, R.~J., {Pepper}, J., {et~al.} 2018, \aj, 156, 102,
  \dodoi{10.3847/1538-3881/aad050}

\bibitem[{{Tauris} \& {Savonije}(1999)}]{1999AATauris}
{Tauris}, T.~M., \& {Savonije}, G.~J. 1999, \aap, 350, 928,
  \dodoi{10.48550/arXiv.astro-ph/9909147}

\bibitem[{{Toonen} {et~al.}(2017){Toonen}, {Hollands}, {G{\"a}nsicke}, \&
  {Boekholt}}]{2017AAToonen}
{Toonen}, S., {Hollands}, M., {G{\"a}nsicke}, B.~T., \& {Boekholt}, T. 2017,
  \aap, 602, A16, \dodoi{10.1051/0004-6361/201629978}

\bibitem[{{Wambsganss}(1998)}]{1998LRRWambsganss}
{Wambsganss}, J. 1998, Living Reviews in Relativity, 1, 12,
  \dodoi{10.12942/lrr-1998-12}

\bibitem[{{Wiktorowicz} {et~al.}(2021){Wiktorowicz}, {Middleton}, {Khan},
  {Ingram}, {Gandhi}, \& {Dickinson}}]{2021MNRASWiktorowicz}
{Wiktorowicz}, G., {Middleton}, M., {Khan}, N., {et~al.} 2021, \mnras, 507,
  374, \dodoi{10.1093/mnras/stab2135}

\bibitem[{{Willems} \& {Kolb}(2004)}]{2004AAwillems}
{Willems}, B., \& {Kolb}, U. 2004, \aap, 419, 1057,
  \dodoi{10.1051/0004-6361:20040085}

\bibitem[{{Witt} \& {Mao}(1994)}]{1994ApJwittmao}
{Witt}, H.~J., \& {Mao}, S. 1994, \apj, 430, 505, \dodoi{10.1086/174426}

\bibitem[{{York} {et~al.}(2000){York}, {Adelman}, {Anderson}, {Anderson},
  {Annis}, {Bahcall}, {Bakken}, {Barkhouser}, {Bastian}, {Berman}, {Boroski},
  {Bracker}, {Briegel}, {Briggs}, {Brinkmann}, {Brunner}, {Burles}, {Carey},
  {Carr}, {Castander}, {Chen}, {Colestock}, {Connolly}, {Crocker}, {Csabai},
  {Czarapata}, {Davis}, {Doi}, {Dombeck}, {Eisenstein}, {Ellman}, {Elms},
  {Evans}, {Fan}, {Federwitz}, {Fiscelli}, {Friedman}, {Frieman}, {Fukugita},
  {Gillespie}, {Gunn}, {Gurbani}, {de Haas}, {Haldeman}, {Harris}, {Hayes},
  {Heckman}, {Hennessy}, {Hindsley}, {Holm}, {Holmgren}, {Huang}, {Hull},
  {Husby}, {Ichikawa}, {Ichikawa}, {Ivezi{\'c}}, {Kent}, {Kim}, {Kinney},
  {Klaene}, {Kleinman}, {Kleinman}, {Knapp}, {Korienek}, {Kron}, {Kunszt},
  {Lamb}, {Lee}, {Leger}, {Limmongkol}, {Lindenmeyer}, {Long}, {Loomis},
  {Loveday}, {Lucinio}, {Lupton}, {MacKinnon}, {Mannery}, {Mantsch}, {Margon},
  {McGehee}, {McKay}, {Meiksin}, {Merelli}, {Monet}, {Munn}, {Narayanan},
  {Nash}, {Neilsen}, {Neswold}, {Newberg}, {Nichol}, {Nicinski}, {Nonino},
  {Okada}, {Okamura}, {Ostriker}, {Owen}, {Pauls}, {Peoples}, {Peterson},
  {Petravick}, {Pier}, {Pope}, {Pordes}, {Prosapio}, {Rechenmacher}, {Quinn},
  {Richards}, {Richmond}, {Rivetta}, {Rockosi}, {Ruthmansdorfer}, {Sandford},
  {Schlegel}, {Schneider}, {Sekiguchi}, {Sergey}, {Shimasaku}, {Siegmund},
  {Smee}, {Smith}, {Snedden}, {Stone}, {Stoughton}, {Strauss}, {Stubbs},
  {SubbaRao}, {Szalay}, {Szapudi}, {Szokoly}, {Thakar}, {Tremonti}, {Tucker},
  {Uomoto}, {Vanden Berk}, {Vogeley}, {Waddell}, {Wang}, {Watanabe},
  {Weinberg}, {Yanny}, {Yasuda}, \& {SDSS Collaboration}}]{2000SDSSyork}
{York}, D.~G., {Adelman}, J., {Anderson}, John~E., J., {et~al.} 2000, \aj, 120,
  1579, \dodoi{10.1086/301513}

\end{thebibliography}
\end{document}